\documentclass{emulateapj}

\def\persqcm{$\rm cm^{-2}$}

\def\htoo{$\rm H_2$}
\def\h2{$\rm H_2$}
\def\error{$\pm$}
\def\e#1{$\times 10^{#1}$}
\def\tenup#1{10$^{#1}$}
\def\asec{\arcsec}

\def\deg{\arcdeg}

\def\kms{km~s$^{-1}$}

\def\solmass{$\rm M_{\sun}$}

\newcommand{\by}{$\times$}

\newcommand{\convunits}{$\rm cm^{-2}\,(K\,km\,s^{-1})^{-1}$}

\newcommand{\jybks}{$\rm Jy\,b^{-1}\,km\,s^{-1}$}
\newcommand{\jykms}{$\rm Jy\,km\,s^{-1}$}
\newcommand{\mjb}{$\rm mJy\:beam^{-1}$}

\newcommand{\oiii}{[\ion{O}{3}]}
\newcommand{\hbeta}{H$\beta$}
\newcommand{\msunsqpc}{\solmass~pc$^{-2}$}
\newcommand{\bit}{\begin{itemize}}
\newcommand{\eit}{\end{itemize}}
\newcommand{\beamsz}[2]{${#1}'' \times {#2}''$}

\begin{document}

\title{Structure and Kinematics of Molecular Disks in Fast-rotator Early-type Galaxies}
\author{Lisa M. Young}
\affil{New Mexico Tech, 801 Leroy Place, Socorro, NM 87801}
\email{lyoung@physics.nmt.edu}
\author{Martin Bureau and Michele Cappellari}
\affil{University of Oxford, Sub-department of Astrophysics,
Denys Wilkinson Bldg., Keble Road, Oxford, OX1 3RH, UK}

\shorttitle{Molecular Disks in Early-type Galaxies}

\slugcomment{ApJ, accepted 23 December 2007}

\begin{abstract}
We present interferometric observations resolving the CO emission in the
four gas-rich
lenticular galaxies NGC~3032, NGC~4150, NGC~4459, and NGC~4526, and
we compare the CO distribution and kinematics to those of the stars and ionized
gas.
Counterrotation documents an external origin for the gas in at least one case
(NGC 3032),
and the comparisons to stellar and ionized gas substructures in all four galaxies
offer insights into their formation histories.
The molecular gas is found in kpc-scale disks with
mostly regular kinematics and average surface densities of 100 to 200
\msunsqpc.  The disks are well aligned with the stellar
photometric and kinematic axes.
In the two more luminous Virgo
Cluster members NGC~4459 and NGC~4526 the molecular gas shows excellent
agreement with circular velocities derived independently from detailed
modeling of stellar kinematic data.
There are also two puzzling instances of disagreements between stellar kinematics and
gas kinematics on sub-kpc scales.  In the inner arcseconds of NGC~3032 the
CO velocities are significantly lower
than the inferred circular velocities, and the reasons may possibly
be related to the external origin of the gas but are not well understood.
In addition, the very young population of stars in the core of NGC~4150
appears to have the opposite sense of rotation from the molecular gas.
\end{abstract}


\section{Introduction}

Early-type galaxies, the ellipticals and lenticulars, are generally poorer in
cold gas than spirals \citep[e.g.][]{lees91} .  This fact is, of course, ultimately responsible
for the color difference between early- and late-type galaxies and their
locations in the red sequence or the blue cloud \citep[e.g.][]{baldry04}.  However, cold atomic and
molecular gas are not entirely absent from all early-type galaxies, and therein
lie some important clues to both the past and the future of early-type
galaxies.  The origin of the cold gas in early-type galaxies can serve as a
tracer of their assembly histories, and the properties of the cold gas (the raw
material for star formation) offer insights into possible morphological
and dynamical evolution through star formation.

\citet{FG76} predicted that mass loss from evolved stars in early-type
galaxies should produce detectable amounts of gas over a Hubble time.  More
recent work \citep[e.g.][]{ciotti91} makes predictions which are different
quantitatively but not qualitatively.  Of course, it is likely that the gas
recycled into the interstellar medium will have a complex thermal history,
including heating to X-ray temperatures and possible cooling all the way to the
formation of molecules \citep{bm96,bm97}.
But since the stellar mass loss is unavoidable and \citet{temi07} have shown
that the dusty ejecta can cool to molecular temperatures in galaxy centers,
it is very natural to
expect that some early-type galaxies could contain cold gas.
In fact, as \citet{temi07} have alluded, the question of why
there are early-type galaxies {\it without} cold gas may be just as
interesting as why there are early-type galaxies with cold gas.

However, if the cold gas ultimately originated in the stars, one might
expect a correlation between the luminosity of an early-type galaxy and
its cold gas content, which has not been seen
\citep{wardle+knapp,H95,lees91,knapp96,combes07}.
The {\it lack} of such a correlation is then often used to suggest that
whatever molecular gas does exist in early-type galaxies is completely
unrelated to the internal stellar mass loss, having been acquired from
external sources such as a satellite galaxy or perhaps even the intergalactic
medium.  In a slightly more sophisticated analysis of both the atomic and
molecular gas contents of lenticular galaxies, \citet{sw06} hypothesized a
variation on this theme in which the atomic gas could be of external origin
whereas the molecular gas could be of internal origin.

Due to the complex nature of the thermal evolution of the stellar
mass loss, plus environmentally dependent interaction with an intracluster medium and
potential time-dependent effects of AGN feedback on the ISM,
we argue that the total gas masses by themselves do not offer compelling
answers about the origin of the cold gas in early-type galaxies.  However,
a comparison of the gaseous and stellar angular momenta should provide
much stronger constraints.  The distribution and kinematics of
the gas can reveal recent gravitational interactions, and gas which is
counterrotating with respect to the stars almost certainly did not originate in those
stars.  Maps which resolve the molecular gas are thus crucial tools which
help us to read the assembly histories of early-type galaxies.

Molecular gas in early-type galaxies is also particularly interesting because
such gas is the raw material for star formation activity.  In recent years
there have been suggestions that current-day star formation is taking place in
as many as 30\% of early-type galaxies at $z=0$ \citep{yi05,kaviraj06}.
Such inferences are at least superficially consistent with the CO detection rates
found by \citet{sw06}, \citet{swy07}, and \citet{combes07}.  But the
distribution of the molecular gas will determine whether the star formation
takes place in a nuclear starburst or in an extended disk and, therefore, it
will also determine the
efficiency with which AGN feedback could disrupt the star formation.
Testing
our theoretical understanding of star formation in early-type galaxies also
requires estimates of the molecular disk sizes and gas surface densities.
Thus, resolved maps of molecular gas hold the keys to understanding
the morphological evolution of early-type galaxies through star formation.

We begin to address these issues on the past and future of early-type
galaxies through interferometric maps which resolve the CO emission in four
nearby lenticular galaxies.
Total molecular masses are commonly available for S0s, but resolved CO maps
are rare \citep[e.g.][]{okuda2005,das05} and the four presented here represent a significant addition to the
literature on the subject.
In all four cases the molecular gas is found in kpc-scale disks, and in one
case the molecular gas counterrotates with respect to the stars.
We also present comparisons between the CO kinematics and circular
velocity curves derived independently from stellar kinematic data.
We discuss the origin of the gas and its implications for mass modeling.
A forthcoming paper will also present comparisons to the stellar populations
in the galaxies and their ionized gas properties, with special emphasis on the
distribution and kinematics of the young stellar population,
showing clear evidence for disk growth through star formation.

\section{Selection and properties of the galaxies}

The four galaxies NGC~3032, NGC~4150, NGC~4459, and NGC~4526 were
observed because of their membership in the {\tt SAURON}
integral-field survey of early-type galaxies \citep[see][]{dezeeuw02},
which means that detailed maps of the stellar kinematics, ionized-gas
kinematics, and stellar populations are available for the regions
within an effective radius
\citep[e.g.][]{emsellem04,sarzi06,kuntschner06}. They also have CO
detections \citep{thronson89,sage+wrobel,lees91,combes07}, making them
suitable for high spatial resolution comparisons of the stellar and
molecular disks. Basic parameters of the galaxies are summarized in
Table \ref{sampletable}.

NGC~3032 is a relatively faint ($M_B=-18.8$) field lenticular with
complex stellar kinematics. Having high angular momentum per unit mass, it belongs to the so-called fast rotator class \citep{emsellem07}, and has a small counter-rotating core ($R\approx2\arcsec$;
\citealt{mcdermid06a}). The ionised gas is co-spatial with dusty
spiral arms extending to about an effective radius $R_{\rm e}$ and is
also counter-rotating \citep{sarzi06}. \hbeta\ emission is the
strongest in the {\tt SAURON} sample while the ratio \oiii/\hbeta\ is
weakest (except in the very center). The absorption linestrength
distributions are unusual, suggesting young stars everywhere with a
lower metallicity and $\alpha$-enhancement in the center
(\citealt{kuntschner06}; Kuntschner et al.\ 08, in prep.), probably
indicating a recent starburst.

NGC~4150 belongs to the Coma~I Cloud and is low-luminosity
($M_B=-18.5$) lenticular in many respects similar to NGC~3032. It is a
fast rotator \citep{emsellem07} with moderate rotation and a dip in
the central velocity dispersion \citep{emsellem04}, and also harbours
a central counter-rotating core \citep{mcdermid06b} where messy dust
is found. Contrary to NGC~3032, however, the extended ionised gas is
co-rotating with the bulk of the stars, although slightly misaligned
\citep{sarzi06}. The linestrengths again indicate young stars
everywhere, particularly in the center, but the metallicity and
$\alpha$ elements are rather flat (\citealt{kuntschner06}; Kuntschner
et al.\ 08, in prep.).

NGC~4459 is an average ($M_B=-20.0$) Virgo lenticular with rapid
rotation and the pinch of the stellar isovelocity contours
characteristic of decoupled central disks \citep{emsellem04}. This
disk is co-spatial with a regular but flocculent dust disk and rapidly
rotating ionised-gas with low \oiii/\hbeta\ ratio (except in the very
center; \citealt{sarzi06}). The linestrengths indicate much younger
stars and higher metallicity in the central disk
(\citealt{kuntschner06}; Kuntschner et al.\ 08, in prep.).

NGC~4526 is a Virgo lenticular similar to NGC~4459 but slightly
brighter ($M_B=-20.7$). It harbours what is probably the strongest
decoupled central stellar disk in the {\tt SAURON} sample (observed
nearly edge-on), embedded in a more slowly rotating
bulge \citep{emsellem04}. The bulge has a slight triaxiality which
suggests the presence of a bar.  A rather large but regular dust and
ionised-gas disk is co-spatial with the stellar disk, and has
uniformly low \oiii/\hbeta\ ratio \citep{sarzi06}. The linestrengths
clearly show that the central disk is much younger and metal rich than
the bulk of the galaxy (\citealt{kuntschner06}; Kuntschner et al.\ 08,
in prep.).

The four galaxies studied here thus fall naturally into two
classes. NGC~3032 and NGC~4150 are relatively faint and have low velocity dispersion, but harbour small counter-rotating cores and have
extended marginally ordered gaseous disks with pervasive young
stars. NGC~4459 and NGC~4526 have rapid rotation but even more rapidly
co-rotating decoupled (and spatially well-constrained) central stellar
disks, co-spatial with regular dust disk populated with young
stars. The former thus point to a more troubled recent past, perhaps
involving accretion or minor mergers, while the latter suggest a more
peaceful immediate past dominated by secular evolution. The current
molecular gas observations should thus allow to build on these data
and test those ideas further.

\begin{deluxetable*}{lcccc}
\tablewidth{6in}
\tablecaption{Sample Galaxies -- Basic Properties
\label{sampletable}}
\tablehead{
\colhead{} & \colhead{NGC 3032} & \colhead{NGC 4150} & \colhead{NGC 4459} &
\colhead{NGC 4526} \\
}
\startdata
RA (J2000.0) & 09 52 08.2 & 12 10 33.6 & 12 29 00.0 & 12 34 03.0\\
Dec & +29 14 10 & +30 24 06 & 13 58 43 & 07 41 57\\
Velocity (\kms) & 1533 (5) & 226 (22) & 1210 (16) & 575 (24)\\
Distance (Mpc) & 21.4 & 13.4 & 15.7 & 16.4 \\
Type & SAB0$^0$(r) & S0$^0$(r)? & S0$^+$(r) & SAB0$^0$(s): \\
$T$  & $-1.7 $ & $-2.4$ & $-2.0$ & $-1.6$ \\
$M_B$ &  $-18.8$ & $-18.5 $ & $-20.0$ & $-20.7$ \\
$(B-V)_e$ & 0.63 & 0.83 & 0.97 & 0.98 \\
$R_e$  (\asec) & 9 & 18 & 35 & 44 \\
$\sigma_0$ (\kms) & 82 & 148 & 174 & 256 \\
CO flux (\jykms)  & 93 (18) & 26 (5) & 56 (11) & 180 (36) \\
H$_2$ mass ($10^8$ \solmass) & 5.0 (1.0) & 0.55 (0.11) & 1.6 (0.3) & 5.7 (1.1) \\
\enddata
\tablecomments{Data are taken from NASA's Extragalactic Database (NED) and the
Lyon Extragalactic Database (LEDA).
The velocities are stellar measurements with the exception of NGC 3032's, which
is a HI velocity.
Distance measurements are from \citet{tonry01} and \citet{mei05} and $M_B$ are taken
from \citet{emsellem07}.
CO fluxes are derived from the present observations, and H$_2$ masses use a
CO-to-H$_2$ conversion factor of 3.0\e{20} \convunits.
}
\end{deluxetable*}

\section{CO data}

NGC 3032, NGC 4150, NGC 4459, and NGC 4526 were observed in the
$^{12}$CO $J=1\rightarrow 0$ line with the 10-element
Berkeley-Illinois-Maryland
Association (BIMA) millimeter interferometer at Hat Creek, CA
\citep{welch96}.
These observations were carried out in the C configuration (projected
baselines 3 to 34 k$\lambda$) in the spring and fall of 2003.
Additional data for NGC 3032 and NGC 4526 were obtained in the B
configuration (projected baselines 4 to 83 k$\lambda$) in March 2003.
Total observing times were 8 to 17 hours per galaxy
in the C configuration and 10 to 16 hours in the B configuration.
A single pointing centered on the optical position was used for all
galaxies.
Each observation covered a velocity range of 1300 \kms, and
these data have sensitivity to structures from point sources up
to objects 60--80\arcsec\ in diameter.
System temperatures were mostly in the 300--500 K range.
Table \ref{obstable} summarizes important parameters of these observations.

\begin{deluxetable*}{lccccc}
\tablewidth{5in}
\tablecaption{CO Observation Parameters
\label{obstable}}
\tablehead{
\colhead{Galaxy} & \colhead{Flux cal} & \colhead{Phase cal} &
\colhead{Velocity Range} & \colhead{FOV}
& \colhead{3mm cont.} \\
\colhead{} & \colhead{} & \colhead{} & \colhead{\kms} & \colhead{kpc} &
\colhead{mJy}
}
\startdata
NGC 3032 & 3c273 & 0927+390, 0854+201 & (876, 2128) & 10.5 & $<$ 8\\
NGC 4150 & 3c273 & 1159+292 & ($-$455, 766) & 6.6 & $<$ 14 \\
NGC 4459 & 3c273 & 3c273    & (547, 1776) & 7.9 & $<$ 11 \\
NGC 4526 & Mars, 3c273 & 3c273 & ($-$233, 1050) & 7.9 & $<$ 14 \\
\enddata
\tablecomments{Field of view (FOV) is the FWHM of the primary beam
(100\asec) at the distances in Table \ref{sampletable}.
}
\end{deluxetable*}

Reduction of the BIMA data was carried out using standard tasks in the
MIRIAD package \citep{sault95}.
Electrical line length calibration was applied to all tracks.
The C configuration data for NGC 3032 and 60\% of the C configuration data
for NGC 4526 were also explicitly corrected for amplitude
decorrelation on longer baselines using data from an atmospheric phase
monitor and  the MIRIAD task {\it uvdecor} \citep{lay99, akeson98, regan01,
wong01}.
The atmospheric decorrelation is estimated using a small interferometer
with a fixed 100 meter baseline which measures the rms path length
difference in the signal from a commercial broadcast satellite.
Data requiring the decorrelation correction had rms path length
differences in the range of 300 to 700 microns, and the median amplitude
correction factor was about 13\%.  However, the longest baselines in these
datasets were later flagged because of their very poor phase stability.
The remainder of the tracks were taken in stabler weather and were not
explicitly corrected for decorrelation, because normal amplitude calibration
can take out most of the effect \citep{wong01}.

Absolute flux calibration was based on observations of Mars and 3C273, and
comparisons of flux measurements on all the observed calibrators suggest that
the absolute flux uncertainties are in the range of 15\%--20\%.
Phase drifts as a function of time were corrected by means of a
nearby calibrator observed every 30 to 40 minutes.
Gain variations as a function of frequency were corrected by the online
passband calibration system.
Inspection of data for 3C273 indicate that residual
passband variations are on the order of 10\% or less in amplitude and
2\deg\ in
phase across the entire band.

The calibrated visibility data were weighted by the inverse square of the
system temperature and the inverse square of the amplitude
decorrelation correction factor (if used), then Fourier transformed.
Dirty images were lightly deconvolved with the Clark clean algorithm
\citep{BGC},
as appropriate for these compact, rather low signal-to-noise detections.
No continuum subtraction was needed.
Integrated intensity and mean velocity maps were produced by the masking
method:
the deconvolved image cube was smoothed along both spatial and velocity
axes, and the smoothed cube was clipped at  $\pm 2.5$ times the rms noise in
a channel.
The clipped version of the smoothed cube was then used as a mask to define a
three-dimensional volume in the original, unsmoothed cube in which the
emission was integrated over velocity
\citep{wong01,regan01}.
Velocity maps were also constructed from Gaussian fits to the line
profile at each position.

Integrated spectra were constructed by first using the integrated intensity
(moment 0) maps to define the spatial extent of the emission, then
integrating over that fixed extent for every channel.
Continuum images were also made by averaging all of the line-free channels in
the final spectral line cubes, but no continuum emission was detected.
Table \ref{obstable} gives 3$\sigma$ limits for point source continuum
emission at the centers of the galaxies, and
Table \ref{imgtable} gives beam size and sensitivity information for the
final spectral line cubes.

\begin{deluxetable}{lccccc}
\tablewidth{0pt}
\tablecaption{CO Image Properties
\label{imgtable}}
\tablehead{
\colhead{Galaxy} &
\colhead{Beam} & \colhead{Beam} &
\colhead{Chan.} & \colhead{noise} & \colhead{N(H$_2$)} \\
\colhead{} & \colhead{\asec} & \colhead{kpc} &
\colhead{\kms} & \colhead{Jy bm$^{-1}$} & \colhead{\tenup{20} cm$^{-2}$}
}
\startdata
NGC3032 & 3.4$\times$2.7 & 0.4$\times$0.3 & 40.4 & 0.013 & 47.7 \\
         & 8.1$\times$6.7 & 0.8$\times$0.7 & 20.2 & 0.026 & 8.0 \\
         & 6.1$\times$5.0 & 0.6$\times$0.5 & 12.12 & 0.025 & 8.3 \\
NGC4526 & 5.0$\times$3.8 & 0.4$\times$0.3 & 20.08 & 0.021 & 18.1 \\
         & 4.3$\times$3.1 & 0.3$\times$0.2 & 40.16 & 0.018 & 45.5 \\
NGC4150 & 8.5$\times$5.1 & 0.6$\times$0.3 & 40.04 & 0.021 & 15.9 \\
         & 8.5$\times$5.1 & 0.6$\times$0.3 & 20.03 & 0.029 & 11.0 \\
NGC4459 & 9.0$\times$5.5 & 0.7$\times$0.4 & 40.31 & 0.014 & 9.4 \\
         & 9.0$\times$5.5 & 0.7$\times$0.4 & 20.14 & 0.020 & 6.7 \\
\enddata
\tablecomments{The N(H$_2$) limit is a sensitivity estimate corresponding
to a 3$\sigma$ signal in one channel. }
\end{deluxetable}

\section{Results}
\subsection{Total CO fluxes}

Comparisons of the total CO fluxes in the BIMA images and in previous single
dish data show that the interferometer usually recovers all of the CO emission.
In the case of NGC 4459, the best single dish flux is from a
survey of the SAURON early-type galaxies made with the IRAM 30m telescope by
\citet{combes07}.
That work finds
54.0 \jykms, with a statistical uncertainty of 2.4 \jykms\ and an absolute
calibration uncertainty of 15\% to 20\%.  The present BIMA
data yield 56 \error~11 \jykms.
An earlier, much noisier detection by \citet{thronson89} found 120 \error~55
\jykms.
Thus, the single dish CO
fluxes are entirely consistent with the flux measured from the
interferometric map.
We adopt here a CO-to-H$_2$ conversion factor of 3.0\e{20} \convunits\ and
distances from \citet{tonry01}, scaled down by 3\% following \citet{mei05}.
At 15.7 Mpc, the \htoo\ mass of NGC 4459
is (1.6 \error~0.3) \e{8} \solmass.

Our BIMA-derived CO flux of NGC 3032 (93 \error~18 \jykms) is also consistent
with most of the
previous single dish measurements.
\citet{thronson89}
measured 85 \error~25 \jykms\ in the 45\asec\ beam of the FCRAO 14m
telescope, and \citet{sage+wrobel} measured 92 \error~10 \jykms\ in the
55\asec\ beam of the NRAO 12m.  However, the more recent measurements of
\citet{combes07} give only 46.0\error~1.5 \jykms\ (again, with typically 15\% to 20\%
absolute calibration uncertainty)
in the 22\asec\ beam of
the IRAM 30m telescope.  Since the BIMA data show emission over a region
at least
25\asec\ in diameter, it is plausible that some emission was missed by the
30m beam.  At a distance of 21.4 Mpc, our flux corresponds to (5.0 \error~
1.0) \e{8} \solmass\ of \htoo.

In contrast, the CO flux we find for NGC~4526 (180 \error~36 \jykms) is significantly larger than
previous single dish measurements.
\citet{lees91} quote a value of 118\error 25 \jykms\ measured by
\citet{sage+wrobel} with the NRAO 12m telescope, and \citet{combes07}
measured 118 \error 4 \jykms\ (statistical) at the IRAM 30m telescope.  Since the CO
images show emission over a region nearly 30\asec\ in diameter,
the 30m telescope may have missed significant flux.
Furthermore, the spectrum of \citet{sage+wrobel} is quite asymmetric, which
may indicate mispointing or an inaccurate spectral baseline (neither of
which plague the interferometric fluxes).
For a distance of 16.4 Mpc, our flux corresponds to a mass of (5.7
\error~1.1) \e{8} \solmass\ of \htoo.
Thus, in NGC~3032, NGC~4459, and NGC~4526
the evidence suggests that the interferometric
maps have recovered all of the CO emission.

There is some inconsistency in the single dish CO fluxes of NGC~4150, which
makes it more difficult to judge whether the interferometric data are missing
flux.
The most recent 30m CO flux of NGC 4150 is 30.2 \error~2.4 (statistical) \error~5
(calibration) \jykms\
from \citet{combes07}, whereas the BIMA data yield 26
\error~5 \jykms.   These two measurements are in good agreement.  \citet{ws03} measured
45\error~2 \error~7 \jykms, also with the IRAM 30m telescope, and this
value may not be inconsistent with the previous two.
However, a flux of 77 \error~14 \jykms\ reported by \citet{leroy05} from the
ARO 12m telescope (the old NRAO 12m telescope) is
significantly higher, which might be due to an absolute calibration
uncertainty,
an underestimated baseline level, or even some extended molecular gas which is sampled
by the 55\asec\ beam of the 12m but not by the 22\asec\ beam of the 30m.
At a distance of 13.4 Mpc, our BIMA flux corresponds to an \htoo\ mass of
(5.5 \error~1.1) \e{7} \solmass.

\subsection{NGC 3032}

\begin{figure}
\includegraphics[scale=0.45,bb=18 144 592 618,clip]{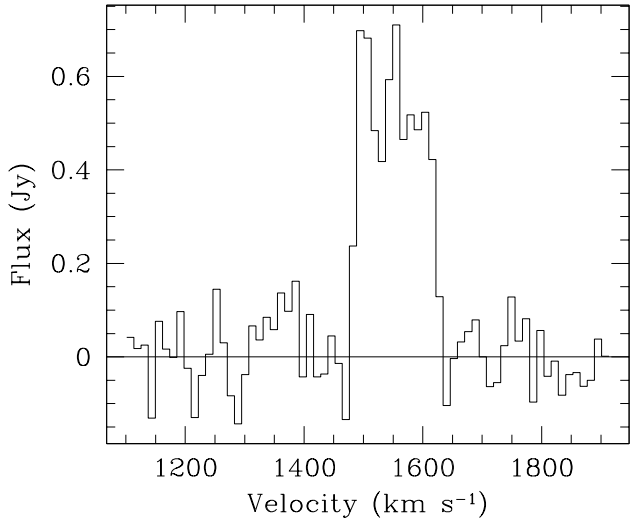}
\epsscale{0.7}
\caption{CO spectrum of NGC 3032.
The spectrum was constructed by first using the integrated intensity
image (Figure \ref{3032stars}) to define an irregular mask region
within which the emission is located.  The intensity was integrated over
the same spatial region for every channel, so the noise in the line-free
regions of the spectrum should be indicative of the noise on the line as
well.
\label{n3032spect}
}
\end{figure}

The CO in NGC 3032 is found in a centrally concentrated structure of rather
narrow linewidth. Emission is detected over 145 \kms\ centered on a systemic
velocity of 1555 \kms\ (Figure \ref{n3032spect}), so
the CO systemic velocity is in good agreement with the stellar absorption
line velocity measurements of 1555~\error~41 \kms\ from \citet{UZC}
and 1559~\error~10 \kms\ from \citet{emsellem04}.
The diameter of the CO emission is approximately 30\asec, with a peak on the
nucleus of the galaxy and a tail extending 15\asec\ to the southeast (Figure
\ref{3032stars}).
Based on a comparison with dust maps (see below), we infer that the molecular gas lies
in a circular disk of radius 14\asec\ (1.5 kpc), yielding an average surface
density of
$\approx 100$ \msunsqpc\ including helium.
NGC 3032 is also known to contain 0.9 \jykms\ (1.0\e{8} \solmass) of HI emission
\citep{duprie96}, giving M(\htoo)/M(HI) $\approx\ 5$.

\begin{figure}
\includegraphics[scale=0.45]{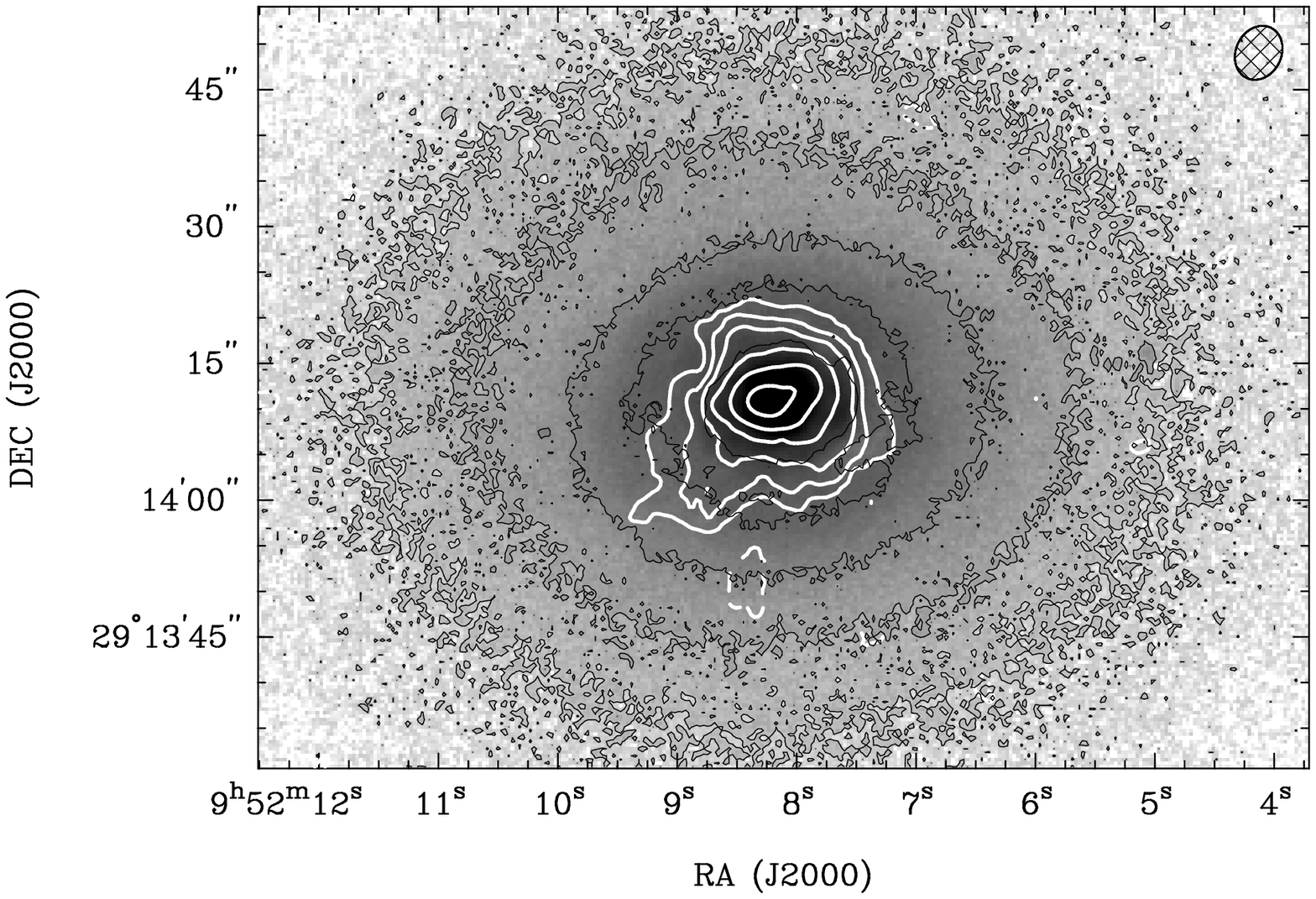}
\epsscale{0.7}
\caption{CO distribution in NGC 3032.
The greyscale and fine black contours show the SDSS $g$ image, with a
contour interval of 0.75 mag (a factor of two).  White contours show the
integrated CO distribution at $-10$, 10, 20, 30, 50, 70, and 90 percent of
the peak (18.1 \jybks = 1.6\e{22} \persqcm\ or 260 \msunsqpc; helium
is not included in the surface mass density).
The hashed ellipse in the upper right corner shows the CO resolution
(\beamsz{6.1}{5.0}).
\label{3032stars}
}
\end{figure}

\begin{figure*}
\includegraphics[scale=0.9]{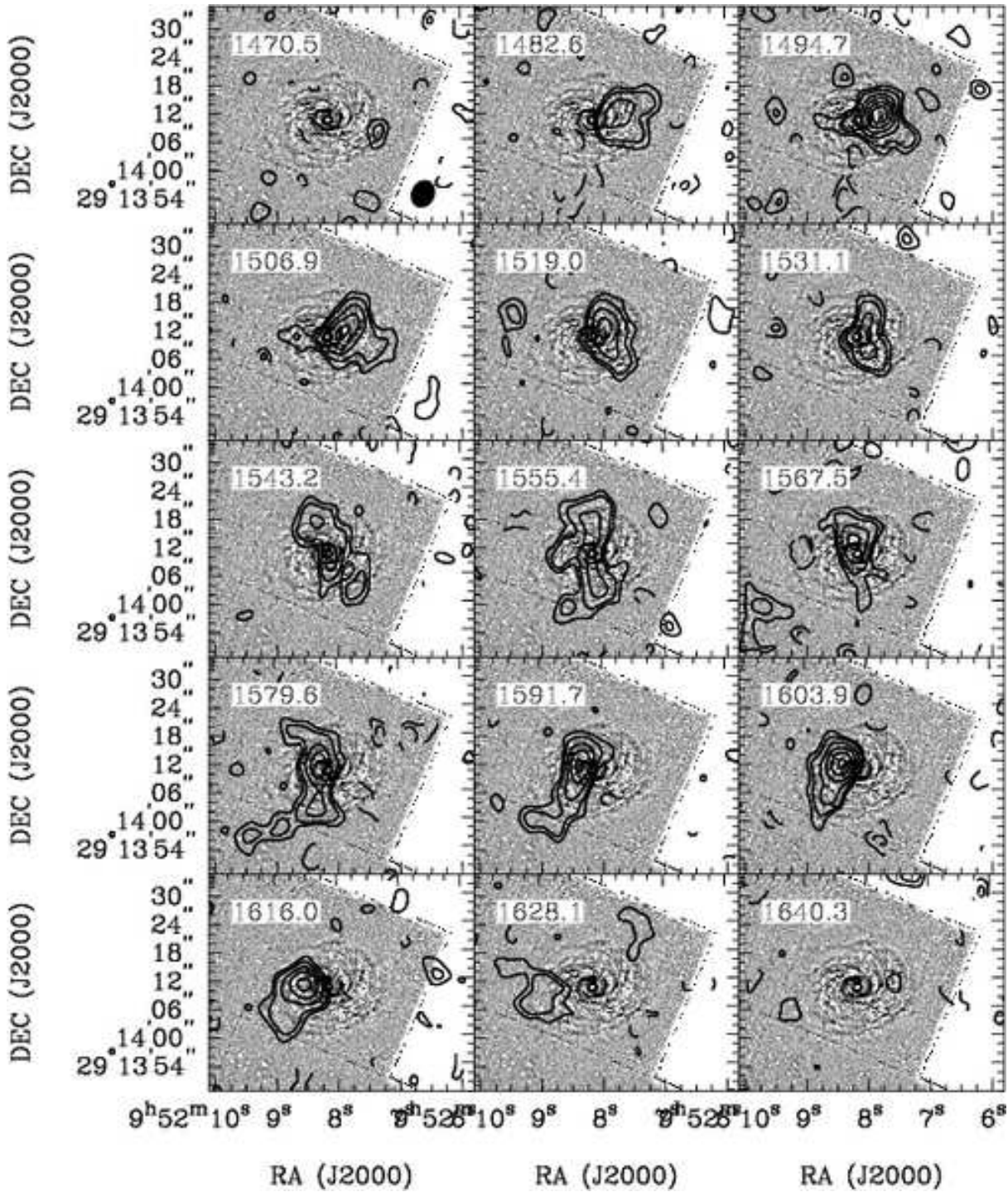}
\caption{CO channel maps for NGC 3032.
The velocity of each channel is indicated in the top left corner and the beam
size in the bottom right corner of the first panel.
Contour levels are $-3$, $-2$, 2, 3, 5, 7, 9, and 11 times the rms noise (25
\mjb).
The greyscale is an unsharp-masked WFPC2 image in the F606W filter.
\label{3032chans}
}
\end{figure*}

Individual CO channel maps (Figure \ref{3032chans}) show good agreement
between the distribution of
molecular gas and dust in NGC 3032; dust and gas are both found in an
inclined, rotating disk.
An unsharp-masked WFPC2 image (Figure
\ref{3032chans}) shows a bright nucleus surrounded by a dark dusty ring (or
two tightly wrapped arms) roughly 6\asec\ in major axis diameter, and beyond
that ring is a disk of flocculent spiral dust features interspersed with
bright point sources.  The dust features cover a region 28\asec\ in diameter.
CO channels have a butterfly-wing structure with compact
emission in the end channels and central channels elongated in the direction
of the kinematic minor axis.
The extent of the CO emission matches that of the dust disk and the
gas kinematic major axis matches the dust morphological major axis.
The southeast ``tail" of emission in Figure \ref{3032stars} is
also visible in the channel maps at 1580 to
1616 \kms\ as emission at about the 75 \mjb\ level, smoothly tracking to
smaller radii with increasing velocity; it does not appear to be associated with a dust feature.

\begin{figure}
\includegraphics[scale=0.5,bb=28 232 567 590,clip]{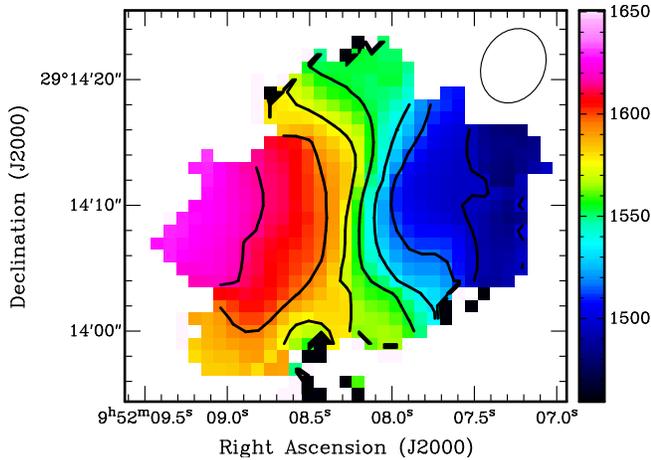}
\epsscale{0.7}
\caption{CO velocity field of NGC 3032.
The velocities are derived from Gaussian fits to the line profile at each
position.  Contours range from 1495 \kms\ to 1615 \kms\ at intervals of 20 \kms.
The beam size is indicated in the top right corner.
\label{n3032vel}
}
\end{figure}

\begin{figure}
\includegraphics[scale=0.5]{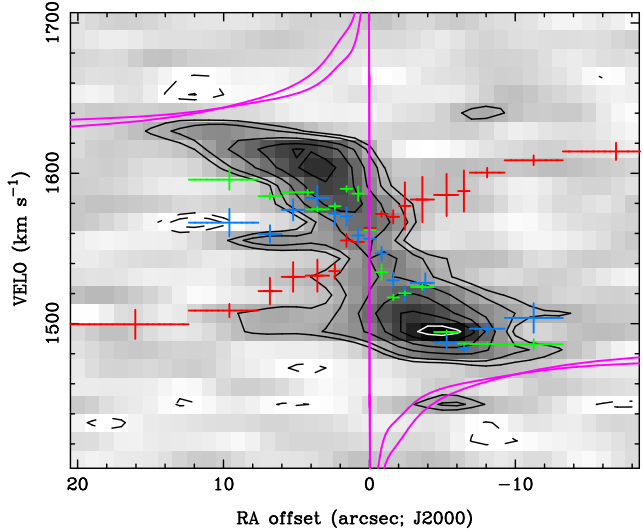}
\caption{Major axis position-velocity diagram for NGC 3032.
The galaxy is sliced through the center at a position angle of 95\deg.
The abcissa is measured along the slice direction with the east side of the
galaxy in positive values on the left side of the plot.
The greyscale and black and white contours show CO intensity (contours are $-3$,
$-2$, 2, 3, 5, 7, 9, and 11 times the rms noise).
Red crosses show the mean stellar velocity from
the SAURON data, green crosses show \hbeta\ velocities and blue show
\oiii.
The two solid magenta lines are two different estimates of the circular
velocity $V_c \sin{i}$ (see Section \ref{ringmodels}).
\label{n3032majax}
}
\end{figure}

Kinematic analysis of the CO in NGC 3032 was made with several different
techniques whose results are all in good agreement.
To the velocity field shown in Figure \ref{n3032vel}
we fitted a model described by an inclined
disk with the rotation curve $V(r) \propto 1 - e^{-r/r_0}$,
as encoded in the National Radio Astronomy Observatory's AIPS task {\it GAL}.
The details of the shape of this rotation curve are not critical, but it has
the desired behavior of rising quickly at small radii and asymptotically
flattening.
We also employed the kinemetry analysis of \citet{davor}, and we
made a tilted ring analysis with the {\it rotcur} task of
the GIPSY package from the Kapteyn Institute, Rijksuniversiteit Groningen.
In using the {\it rotcur} task we followed the procedure outlined by
\citet{swaters99}.

The fitted kinematic center positions are within an
arcsecond of the optical nucleus measured from the SDSS $g$ image.
The systemic velocity of the galaxy is also robustly fitted to be 1555~\error~1
\kms.
The inclination of the disk is poorly constrained by the CO data, however, and is more
accurately taken from the axis ratio of the dust disk which gives
44\deg~\error~4\deg.  The Jeans dynamical modeling we did in this work
is consistent and suggests $i = 42\deg$.

Measurements of the globally averaged CO kinematic position angle fall in the
range 90\deg\ to
97\deg, measured to the receding major axis, while the tilted ring model and the kinemetry both suggest a gentle twist
from $\approx$ 81\deg\ in the central resolution element to
97\deg\ at
8\asec $\lesssim r \le$ 14\asec.  This twist is also suggested by the hint of an
integral-sign shape in the zero velocity curve of the velocity field,
although warping
is not obvious in the images of the dust disk.
A major axis position-velocity diagram sliced through the data cube at 95\deg\ is
shown in Figure \ref{n3032majax}.
This position-velocity diagram shows the typical pattern with a steeply
rising inner portion followed by a flattening (especially noticeable on the
eastern side of the galaxy) beyond radii $\approx$ 4\asec.

The kinematic position angle for the molecular disk is consistent with the
photometric position angle, defined as that of the ellipse of inertia of the surface brightness
\citep*[99.6\deg,][]{cappellari07}, but is 180\deg\ offset from the stellar
kinematic angle of $-89^\circ$
\citep{emsellem04,cappellari07}.\footnote{\citet{cappellari07} quote a
stellar kinematic angle of ${\rm PA_{kin}}=91\deg$, as they define it as it as the direction along which $|V|$ is maximum. The velocity fields shown in
\citet{emsellem04} make it clear that this is to the
stellar major axis on the approaching side rather than the receding side.}
Thus, the molecular gas in NGC~3032 is counterrotating with respect
to the bulk of the stars in the galaxy.
However, \citet{mcdermid06a,mcdermid06b} found a large radial age gradient in
the galaxy and a counterrotating stellar core, which suggests recent star
formation in the molecular disk.  The sense of rotation of the molecular gas
is consistent with that of the young counterrotating stars but inconsistent
with that of the older, more extended population.

\subsection{NGC 4150}\label{4150results}

\begin{figure}
\includegraphics[scale=0.45,bb=18 144 592 618,clip]{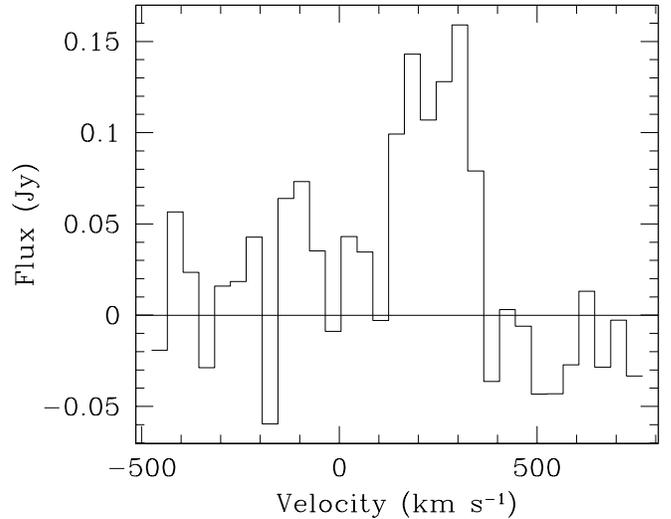}
\caption{CO spectrum of NGC 4150.
\label{n4150spect}
}
\end{figure}

\begin{figure}
\includegraphics[scale=0.45]{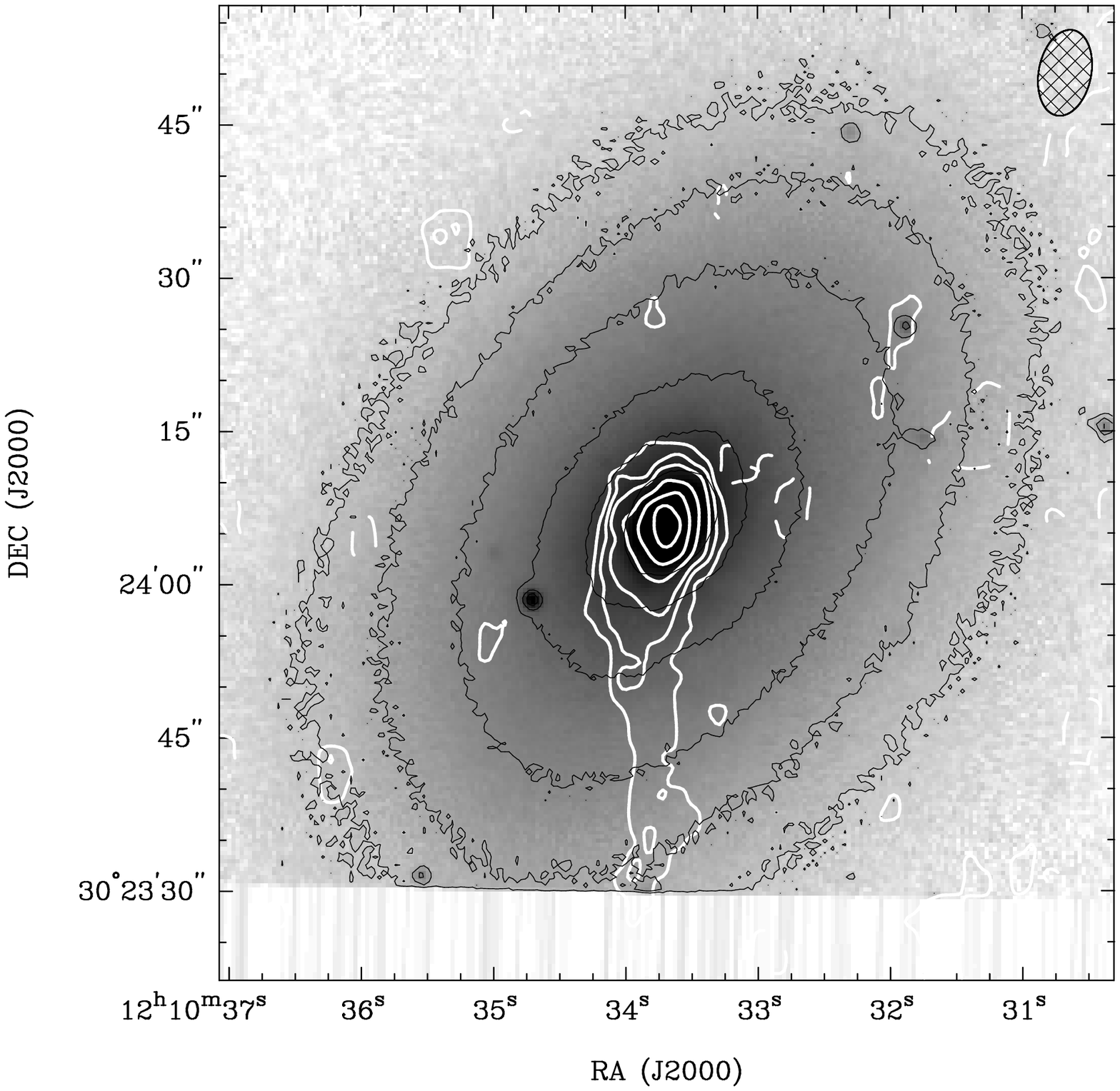}
\caption{CO distribution in NGC 4150.
The greyscale and fine black contours show the SDSS $g$ image, with a
contour interval of 0.75 mag (a factor of two).  White contours show the
integrated CO distribution at $-10$, 10, 20, 30, 50, 70, and 90 percent of
the peak (17.0 \jybks = 1.1\e{22} \persqcm\ or 170 \msunsqpc; helium
is not included in the surface mass density).
The hashed ellipse in the upper right corner shows the CO resolution
(\beamsz{8.5}{5.1}).
\label{n4150mom0}
}
\end{figure}

NGC 4150 shows a relatively small amount of molecular gas in a very compact
structure at the center of the galaxy.
We fit the integrated spectrum in Figure \ref{n4150spect} with a Gaussian
whose central velocity is 239~\error~20 \kms, which we take as the
systemic velocity of the CO emission.
This CO velocity is in good agreement with the stellar velocity
measurements of 226~\error~22 \kms\ \citep{fisher95}, 208~\error~30
\kms\ \citep{UZC}, and 219~\error~10 \kms\ \citep{emsellem04}.
CO emission is detected over a velocity range of 180 \kms, and
the bulk of this gas is within a few arcseconds of the
nucleus of the galaxy (Figure \ref{n4150mom0}).

\begin{figure*}
\includegraphics[scale=0.9]{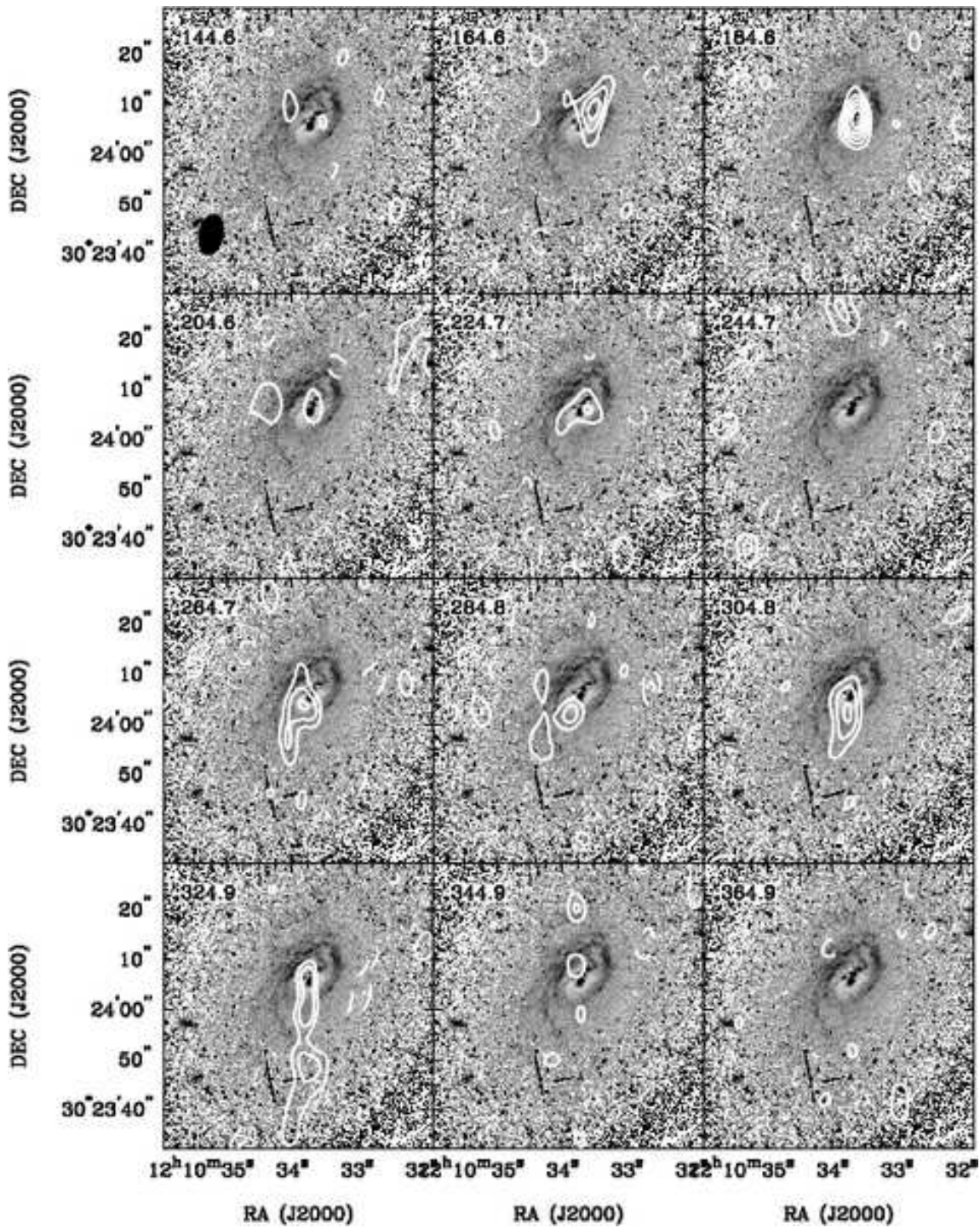}
\caption{CO channel maps for NGC 4150.
The velocity of each channel is indicated in the top left corner and the beam
size is in the bottom left corner of the first panel.
Contour levels are $-3$, $-2$, 2, 3, 4, 5, and 6 times 29 \mjb\ ($\approx
1\sigma)$.
The greyscale is an F606W $-$ F814W ($\approx R-I$) image constructed
from HST WFPC2 data.
\label{n4150chans}
}
\end{figure*}

Several dust structures are apparent in the center of NGC 4150 (Figure
\ref{n4150chans}).   A dark dust lane bisects the nucleus from northwest to
southeast, and the galaxy has more dust clouds a few arcseconds
northwest and east of the nucleus.  There is an irregular dust ring,
stronger in the north, of semimajor and semiminor axes 7.5\asec~\by~4\asec,
with a fainter dust arm curving around the southeast side of
the nucleus to radii $\approx$\ 12\asec.
The peaks in the channel maps suggest that most of the molecular gas is
located in the bisecting dust lane,
and the CO kinematic major axis roughly matches the
major axis of the dust ring.
Figure \ref{n4150mom0} shows a faint tail of emission stretching to 30\asec\
south of the nucleus, also visible in the channel maps from 265
\kms\ to 325 \kms.
\citet{morganti06} show that the 2.5\e{6} \solmass\ of HI in the galaxy
(M(\htoo)/M(HI) = 23) is elongated roughly along the optical major axis
in a structure of  1\arcmin\ diameter, with additional HI to the south and
southwest, so the similarities between the CO and
HI lend credence to the reality of the CO tail.

\begin{figure}
\includegraphics[scale=0.5,bb=28 162 567 680,clip]{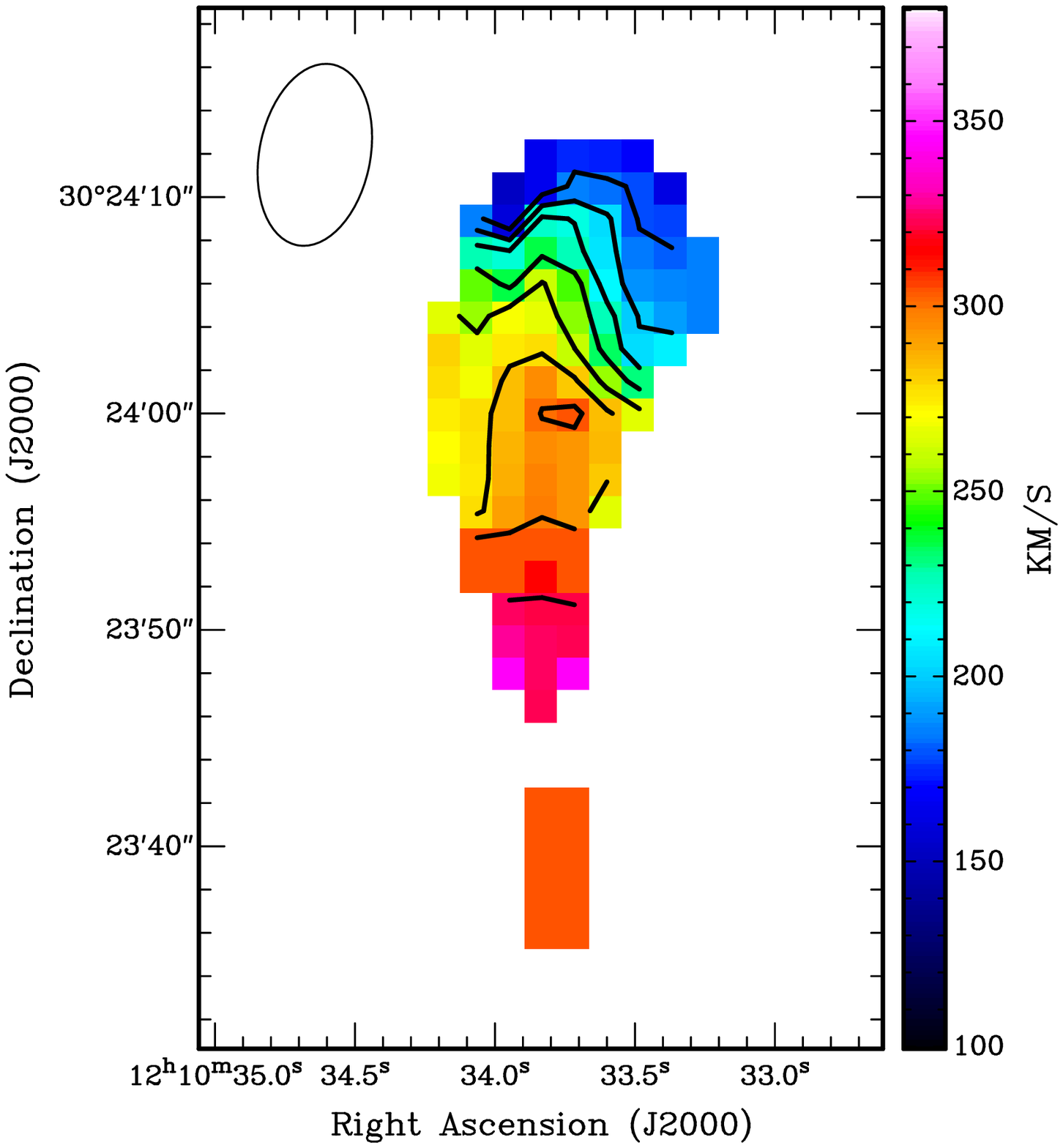}
\epsscale{0.7}
\caption{CO velocity field of NGC 4150.  The velocities shown here are the
intensity-weighted mean velocity (moment 1) rather than Gaussian fits, as the
fits to individual profiles are not robust in this rather low signal-to-noise detection.
Contours are from 180 \kms\ to 320 \kms\ at intervals of 20 \kms.
The beam size is indicated in the top left corner.
\label{n4150vel}
}
\end{figure}

\begin{figure}
\includegraphics[scale=0.5]{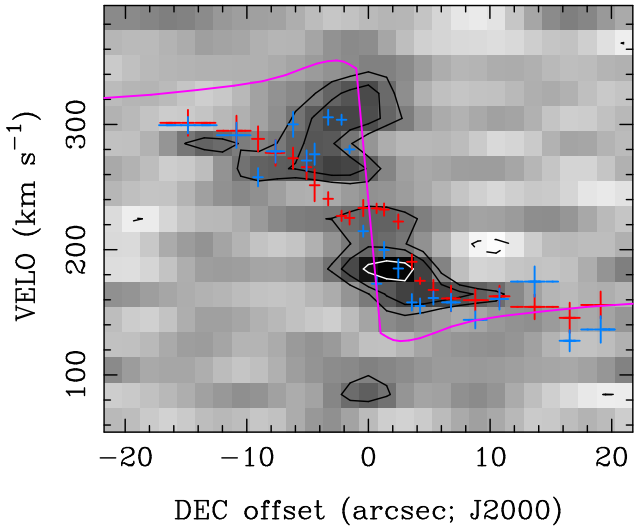}
\caption{Major axis position-velocity diagram for NGC 4150.
The galaxy is sliced through the center at a position angle of $-34$\deg.
The abcissa is measured along the slice with the southern side of the
galaxy in negative values.
The greyscale, black and white contours show CO intensity (contours are $-2$,
2, 3, and 5 times the rms noise level).
Red crosses show the mean stellar velocities from
the SAURON data and blue crosses show \oiii.  (Velocities of \hbeta\ are
not fitted independently of \oiii\ in this galaxy.)
The solid magenta line is $V_c \sin{i}$.
\label{n4150majax}
}
\end{figure}

Due to the small angular size of the CO emission the velocity field
(Figure \ref{n4150vel}) does not offer good constraints on the kinematic
parameters of the galaxy.
\citet{cappellari07} quote both the photometric and stellar kinematic position
angles as $-33^\circ$, with uncertainties of a few degrees.
The orientation of the dust ring (Figure \ref{n4150chans}) is $-34^\circ \pm
5^\circ$, and although the CO velocity field is complex, it is not
inconsistent with this value.
A slice at position angle $-34^\circ$ gives the major axis position-velocity diagram
of Figure \ref{n4150majax}.
It appears that very little of the molecular gas is located beyond the
turnover point in the rotation curve.
The inclination of the galaxy is measured from the axis ratio of the dust ring
to be 54\deg \error 5\deg, consistent with the value of 52\deg\ from the
dynamical modelling \citep{cappellari06}.

There is a counterrotating stellar core in the central arcseconds of NGC~4150
\citep[][and it is also visible in the position-velocity diagram of Figure
\ref{n4150majax}]{mcdermid06a, mcdermid06b}.  This counterrotating core is believed to have formed
recently as the galaxy shows a very strong radial age gradient.
Curiously, though, the core does not appear to have a counterpart
in the molecular gas at the current resolution and sensitivity.
The counterrotating core is also not obvious in the \oiii\ velocities of
NGC~4150 (Figure \ref{n4150majax}), and indeed the CO velocities of this
galaxy are a better match to the \oiii\ velocities than they are to the
stellar velocities.   A good match between CO velocities and ionized gas
velocities suggests the possibility that the ionized gas could trace star
formation activity, but the relationship to the counterrotating stellar
core is not yet clear.

\subsection{NGC 4459}

\begin{figure}
\includegraphics[scale=0.45,bb=18 144 592 618,clip]{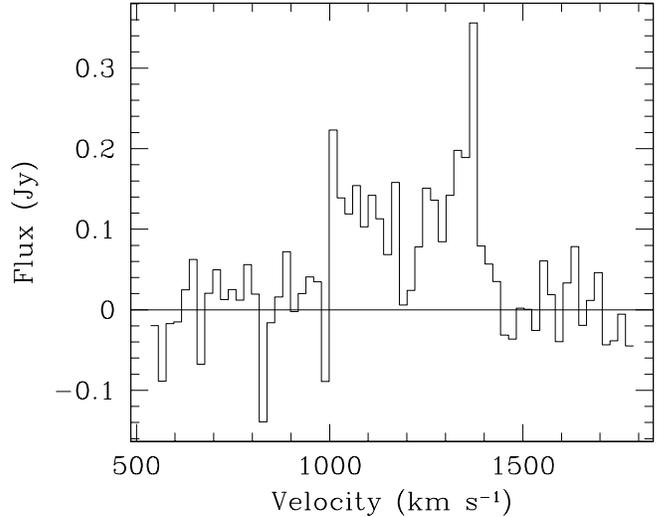}
\caption{CO spectrum of NGC 4459.
\label{n4459spect}
}
\end{figure}

The CO emission from NGC 4459 is also in a central disk, with
hints of a double-horned structure in the integrated spectrum
(Figure \ref{n4459spect}).
A double-horned spectrum is
typically produced by gas in the flat part of the rotation curve.
Emission is detected over a velocity range of 400 \kms\ centered near a systemic
velocity of 1210~\error~20~\kms\  (but see below).
The CO velocity is thus in good agreement with the stellar velocity of
1232~\error~40~\kms\ from \citet{UZC} and 1200~\error~10 \kms\ from
\citet{emsellem04}.

\begin{figure}
\includegraphics[scale=0.45]{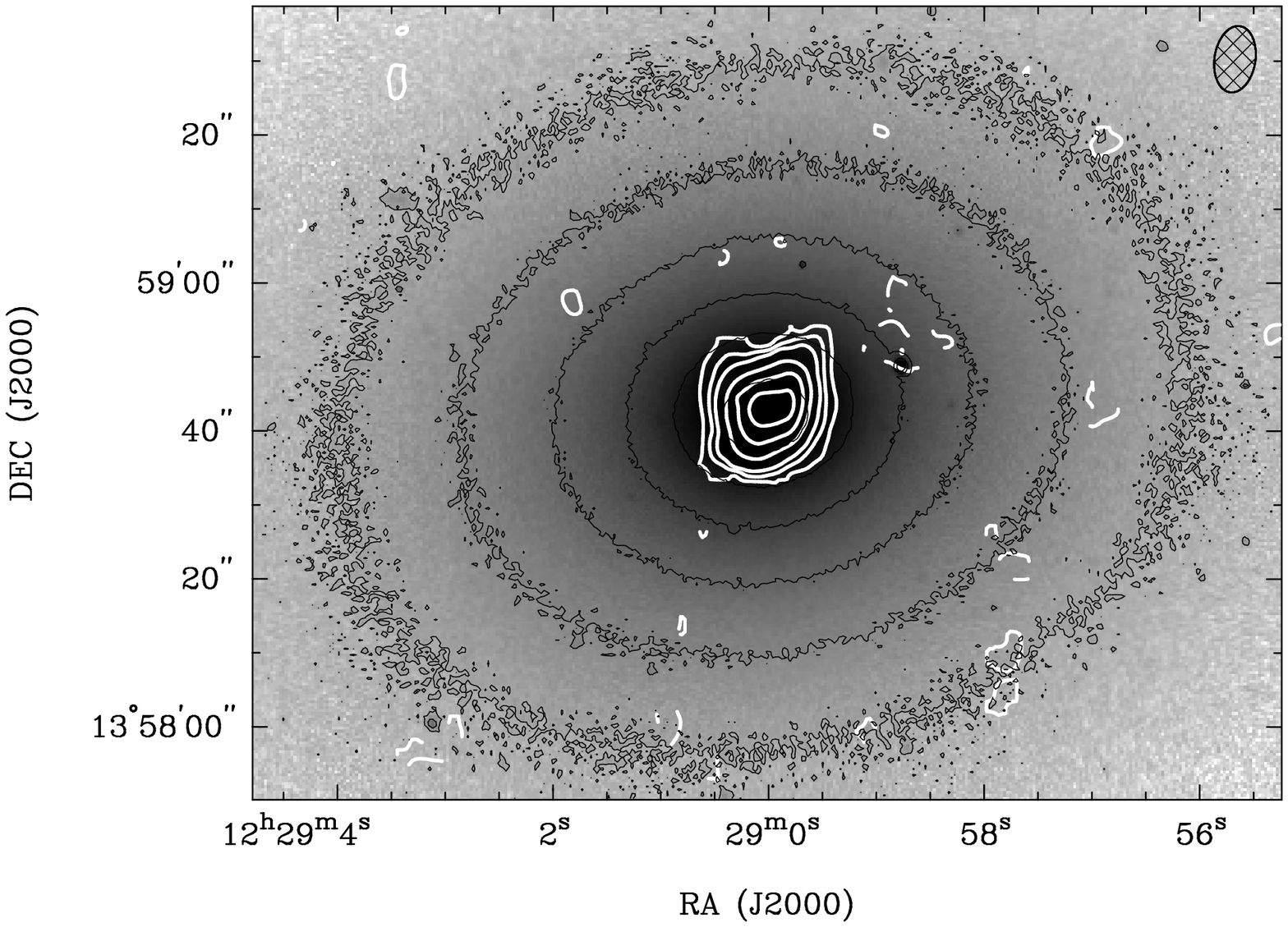}
\caption{CO distribution in NGC 4459.
The greyscale and fine black contours show the SDSS $g$ image, with a
contour interval of 0.75 mag (a factor of two).  White contours show the
integrated CO distribution at $-10$, 10, 20, 30, 50, 70, and 90 percent of
the peak (21.4 \jybks = 1.2\e{22} \persqcm\ or 190 \msunsqpc).
The hashed ellipse in the upper right corner shows the CO resolution
(\beamsz{9.0}{5.5}).
\label{n4459mom0}
}
\end{figure}

\begin{figure*}
\includegraphics[scale=0.9]{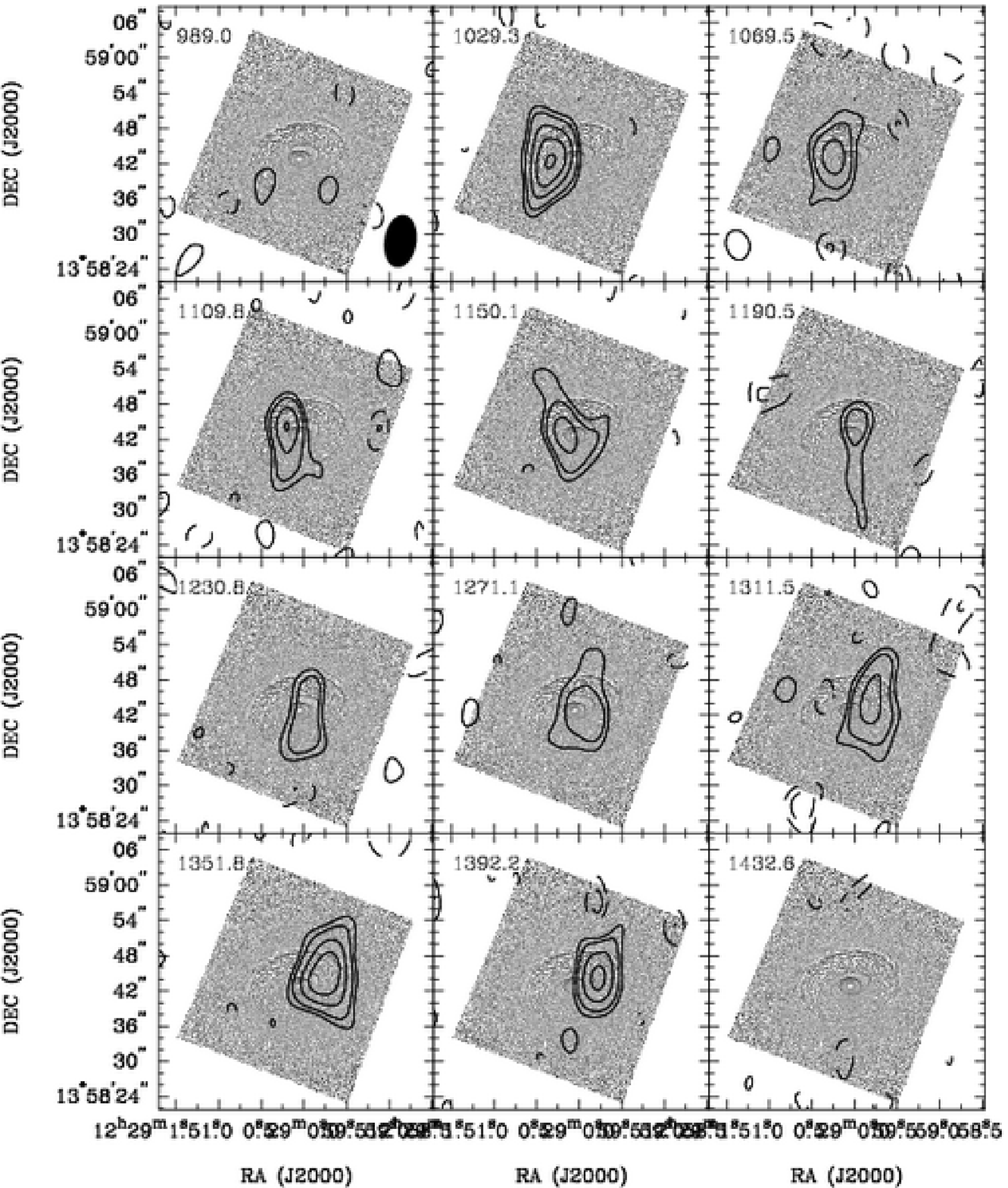}
\caption{CO channel maps for NGC 4459.
The velocity of each channel is indicated in the top left corner and the beam
size in the bottom right corner of the first panel.
Contour levels are $-3$, $-2$, 2, 3, 5, 7, and 9 times  14  \mjb\ ($\approx\
1\sigma)$.
The greyscale is the inner portion of an unsharp masked F555W image from WFPC2.
\label{n4459chans}
}
\end{figure*}

The integrated intensity map (Figure \ref{n4459mom0}) shows a compact
structure centered on the galaxy nucleus, no more than 20\asec\ in diameter.
Little structure is evident, and there is no strong evidence for arms or
tails of molecular gas.
The individual channel maps in Figure \ref{n4459chans} again show a typical disk
pattern with higher intensities at the extreme velocities than near the systemic
velocity.  The centroids of
the emission in each channel show excellent agreement with the extent of the
prominent, well-developed dust disk.  The dust fills a flocculent disk with a
rather sharp outer edge at a semimajor axis of 8.5\asec\ and only a few faint
spiral dust features beyond.  From a radius of about 3.75\asec, two
prominent dusty spiral arms can be traced inward to an inner ring of semimajor axis
2.0\asec.
The larger disk has a major axis position angle of 102\deg\ but the inner
ring appears to have a slightly different orientation, with position angle
$\approx$\ 90\deg.  The bulk of the CO emission appears associated with the
larger-scale, 8.5\asec\ disk.
For a circular disk of radius 8.5\asec\ (670 pc), the average gas surface density is
170 \msunsqpc, including helium.
The north side of the disk must be the near side, suggesting that both the
flocculent and inner spiral arms are trailing.

\begin{figure}
\includegraphics[scale=0.5,bb=28 212 567 590,clip]{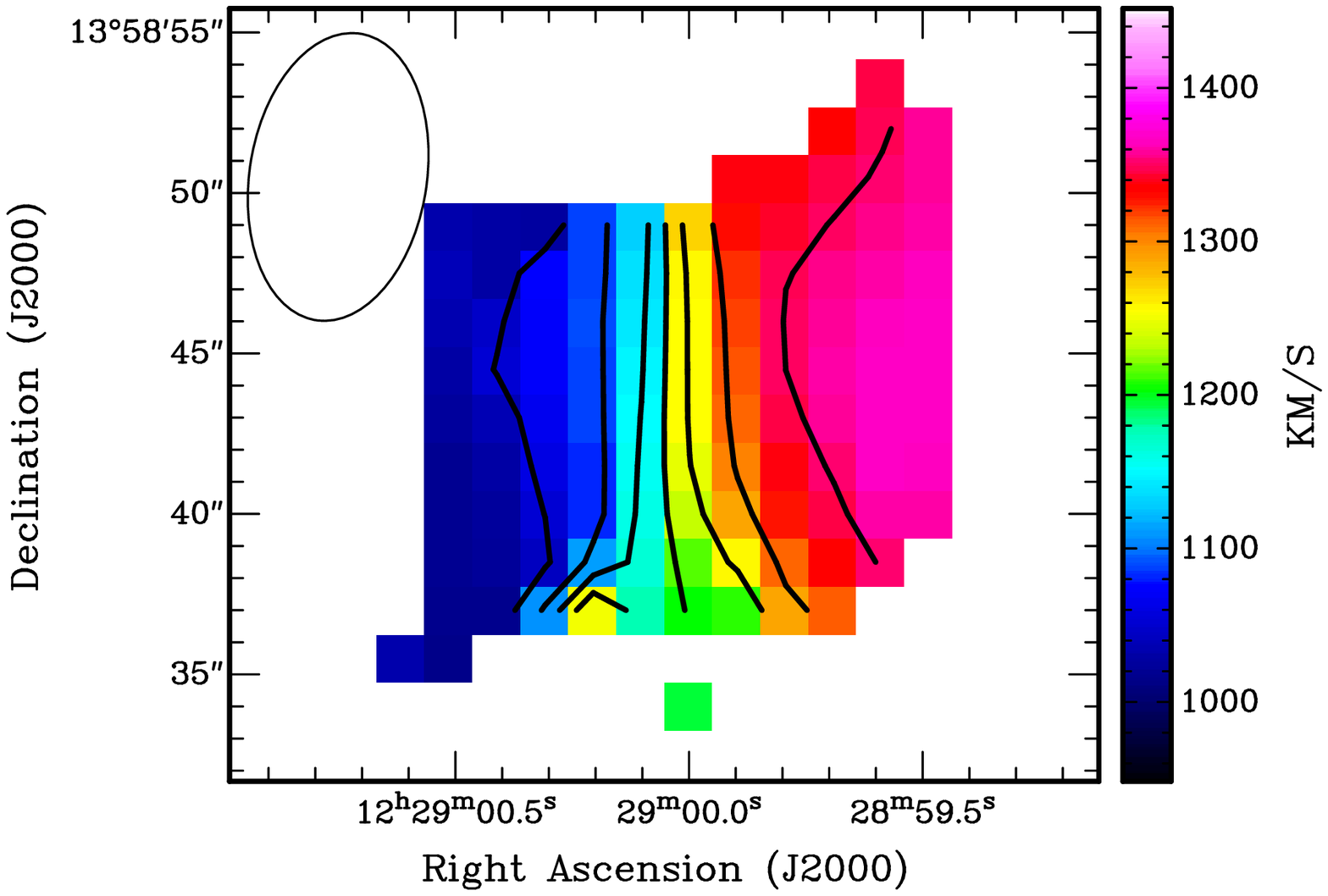}
\caption{CO velocity field of NGC 4459.  Velocities are derived from Gaussian
fits to the line profile at each position.
Contours are from  1050   \kms\ to  1350   \kms\ at intervals of 50   \kms.
The beam size is indicated in the top left corner.
\label{n4459vel}
}
\end{figure}

\begin{figure}
\includegraphics[scale=0.6]{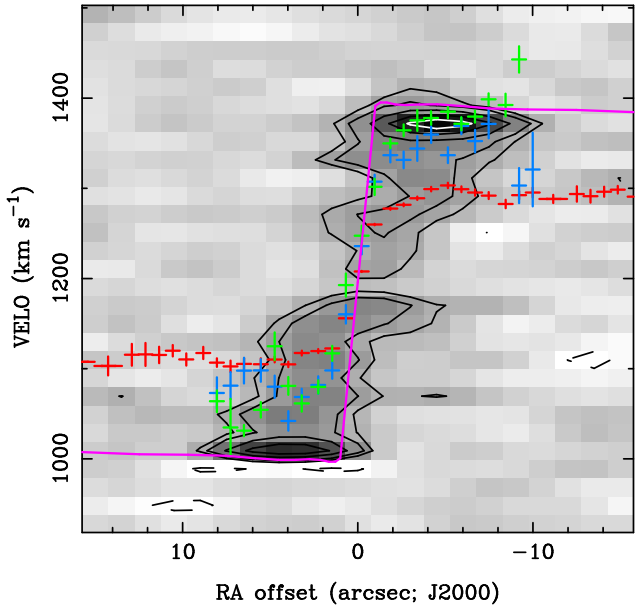}
\caption{Major axis position-velocity diagram for NGC 4459.
The galaxy is sliced through the center at a position angle of $90$\deg.
The east side of the galaxy is at the left side of the figure.
The greyscale, black, and white contours show CO intensity (contours are $-3$, $-2$,
2, 3, 5, 7, and 9 times the rms noise level).
Red symbols show the SAURON stellar velocities along the major axis; blue
symbols show \oiii\ velocities and green symbols are \hbeta.
The solid magenta line is $V_c\:\sin{i}$.
\label{n4459majax}
}
\end{figure}

The velocity field (Figure \ref{n4459vel}) shows predominantly straight,
parallel isovelocity contours, so again the inclination of the gas disk is
better obtained from the dust disk than from the gas itself.
The axial ratio of the dust disk gives
an estimated inclination of $46^\circ\pm 2^\circ$, consistent with 47\deg\ from
dynamical modelling \citep{cappellari06}.
For the kinematic analysis we also
constrain the kinematic center of the galaxy to be the position of
the optical nucleus. 
Stellar isophotes have a position angle of 280\deg\ outside the dust disk,
while a
kinemetric analysis of the stellar velocities
gives position angles of 280\deg\ outside the dust disk, decreasing to 274\deg\
inside it \citep{davor}.  A global kinematic position angle measurement
for the molecular gas,
measured both as in \citet{davor} and by fitting a model velocity field,
yields
values in the range 269\deg\ to 273\deg\ to the receding major axis.  The
molecular gas kinematics are thus very well aligned with the stellar kinematics
and the stellar isophotes.
A tilted ring analysis suggests the systemic
velocity to be 1196\error 2 \kms.
The major axis position-velocity diagram (Figure \ref{n4459majax}) is very
strongly peaked at its maximum velocities, suggesting that the bulk of the
molecular gas is located in the flat part of the galaxy's rotation curve.

\subsection{NGC 4526}

\begin{figure}
\includegraphics[scale=0.45,bb=18 144 592 618,clip]{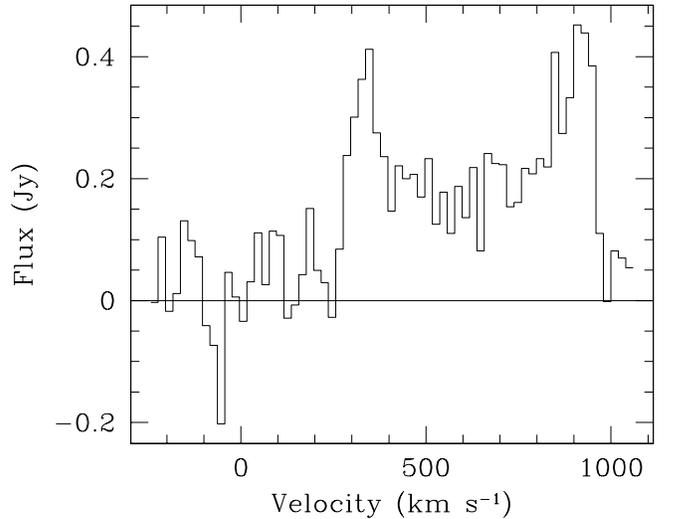}
\caption{CO spectrum of NGC 4526.
\label{n4526spect}
}
\end{figure}

In NGC 4526 the CO lies in a well-developed, nearly edge-on disk.  The
integrated CO profile shows the familiar double-horned shape (Figure
\ref{n4526spect}), with both horns approximately equally bright.  The systemic
velocity of the galaxy is 613 \error\ 10 \kms\ and emission is detected
over a total width of 683 \kms.
A line width of this magnitude, while not unknown, is unusually large for
lenticulars.  In fact, we argue in section \ref{ringmodels} that the galaxy's
maximum circular velocity is about 355 \kms\ and in the compilation of 243
galaxies (54 lenticulars) made by \citet{courteau07}, NGC 4526's circular
velocity is exceeded by only two other lenticulars, one Sa galaxy, and 10
ellipticals.

The observed CO systemic velocity is consistent
with the stellar velocity of 592 \error\ 48 \kms\ given by \citet{UZC}
and 626~\error~10 \kms\ given by \citet{emsellem04},
but is inconsistent with the HI velocity of 448 \error\ 8 \kms\
\citep{RC3,DL73}.  In fact, a comparison of Figure \ref{n4526spect} with the
HI spectrum in \citet{DL73} shows that the latter authors only detected one
``horn", so their HI velocity is not a good estimate of the systemic
velocity.  Interferometric HI observations would undoubtedly be useful in
revealing the relationships between atomic and molecular gas.

\begin{figure}
\includegraphics[scale=0.45]{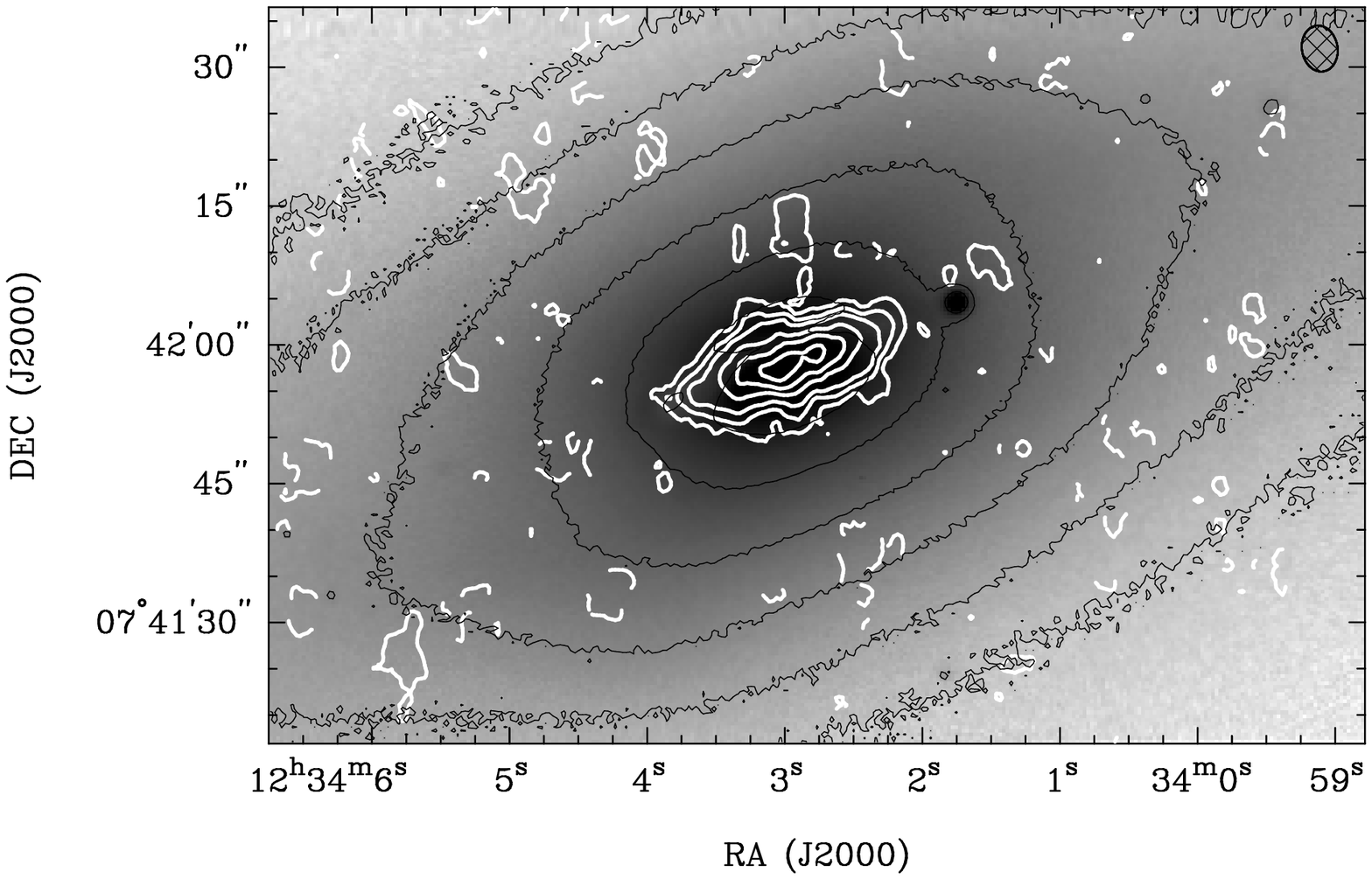}
\caption{CO distribution in NGC 4526.
The greyscale and fine black contours show the SDSS $g$ image with a
contour interval of 0.75 mag (a factor of two).  White contours show the
integrated CO distribution at $-10$, 10, 20, 30, 50, 70, and 90 percent of
the peak (37.1 \jybks = 5.3\e{22} \persqcm\ or 850 \msunsqpc).
The hashed ellipse in the upper right corner shows the CO resolution
(\beamsz{5.0}{3.9}).
A slightly higher resolution integrated intensity image has a peak of
32.2 \jybks\ = 6.8\e{22} \persqcm\ or 1080 \msunsqpc\ at a beam size of 
\beamsz{4.3}{3.1}.
\label{n4526mom0}
}
\end{figure}

The CO distribution (Figure \ref{n4526mom0}) is elongated in the direction
of the optical major axis.  Fitting a two-dimensional Gaussian to the
integrated intensity image reveals that the disk is poorly resolved in the
short dimension; the deconvolved minor axis FWHM is 4.3\asec,
comparable to the beam size.  The position angle of this gas disk is
estimated as $-72.7^\circ$~\error~1.8\deg. 

\begin{figure*}
\includegraphics[scale=0.85]{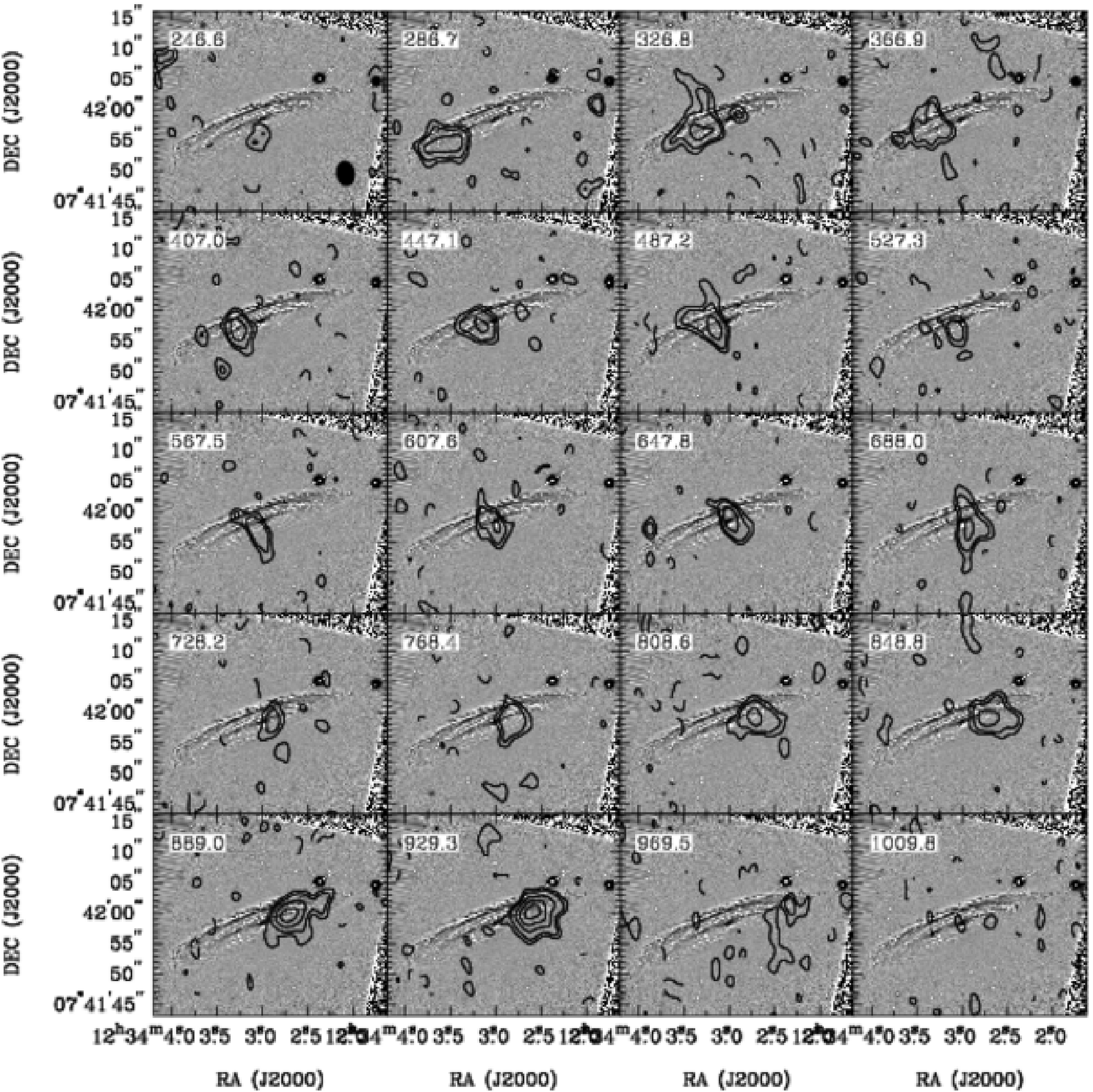}
\caption{CO channel maps for NGC 4526.
The velocity of each channel is indicated in the top left corner and the beam
size in the bottom right corner of the first panel.
Contour levels are $-3$, $-2$, 2, 3, 5, 7, and 9 times 19 \mjb\ ($\approx
1\sigma)$.
The greyscale is an unsharp-masked F555W image from WFPC2.
\label{n4526chans}
}
\end{figure*}

Individual channel maps (Figure \ref{n4526chans}) show a close
correspondence between the CO and dust disks.  Unsharp-masked HST images show
that the dust is confined to radii less than $\approx$ 14\asec\ -- 15\asec.
Although the inclination is quite high, there appear to be at least two
dominant dust rings of radii $\approx$ 14\asec\ and $\approx$ 10\asec.  (These
radii are uncertain because the features are most prominent along the
minor axis.)  The centroids of CO emission in the extreme channels suggest
that the bulk of the CO may be associated with the inner of these two rings.
Assuming the gas to be in a circular disk of radius 14\asec\ (1.1 kpc), the
average surface density is 200 \msunsqpc\ including helium.

\begin{figure}
\includegraphics[scale=0.5,bb=28 262 567 550,clip]{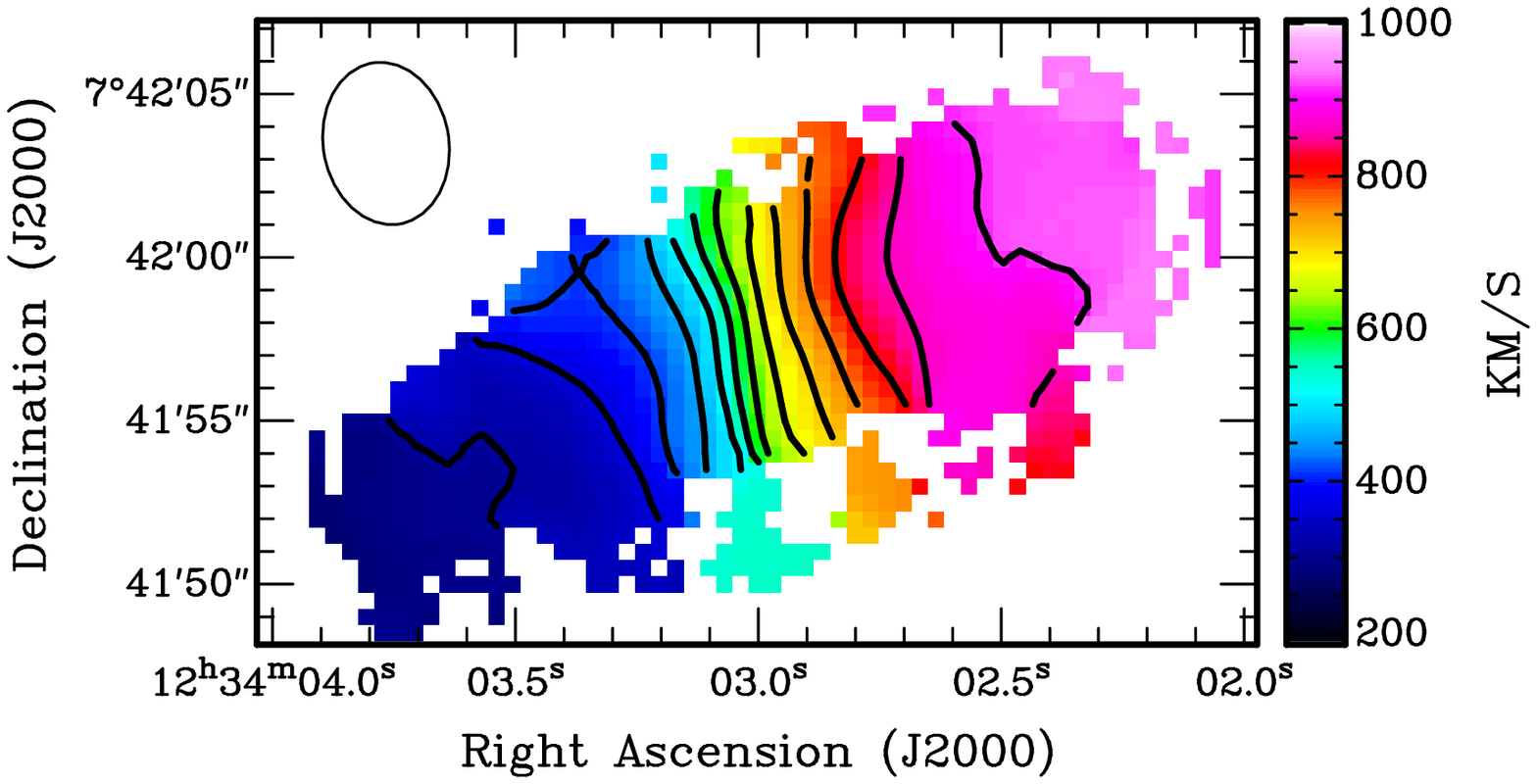}
\caption{CO velocity field of NGC 4526.
The velocities are derived from Gaussian fits to the line profile at each
position.  Contours are from 310 \kms\ to 910 \kms\ at intervals of 50 \kms.
The beam size is indicated in the top left corner.
\label{n4526vel}
}
\end{figure}

\begin{figure}
\includegraphics[scale=0.6]{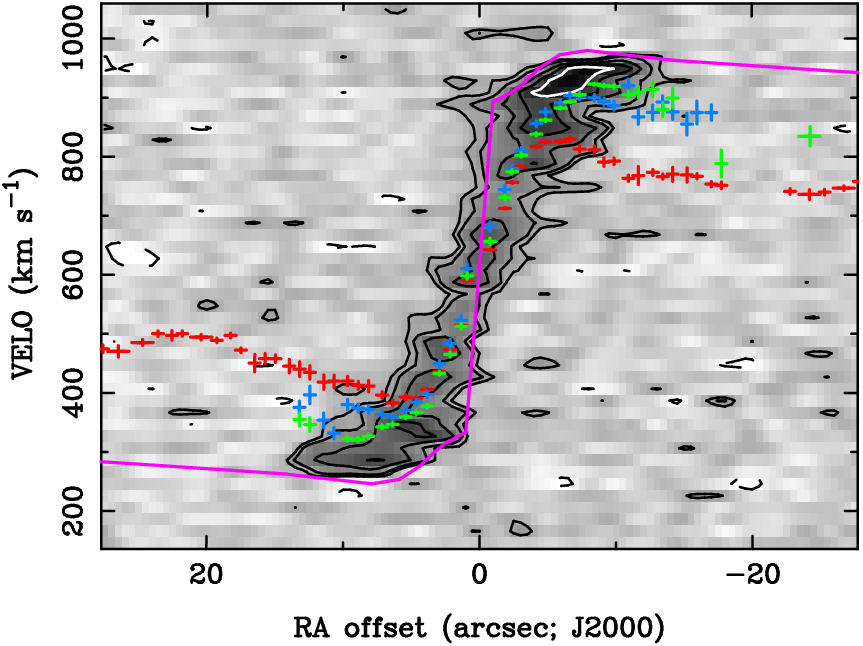}
\caption{Major axis position-velocity diagram for NGC 4526.
The galaxy is sliced through the center at a position angle of $111$\deg.
The east side of the galaxy is at the left side of the figure.
The greyscale, black, and white contours show CO intensity (contours are $-2$,
2, 3, 5, 7, and 9 times the rms noise level).
Red symbols show the SAURON stellar velocities along the major axis; blue
symbols show \oiii\ velocities and green symbols are \hbeta.
The solid magenta line is $V_c\:\sin{i}$.
\label{n4526majax}
}
\end{figure}

As in the case of NGC 4459, the velocity field (Figure \ref{n4526vel})
shows straight, parallel isovelocity contours throughout; it constrains the kinematic
position angle but not the inclination of the gas disk.
However, the dynamical modelling of \citet{cappellari06} gives an inclination
of 79\deg\ which is consistent with values of 75\deg~\error~2\deg\ inferred from
the axis ratio of the dust disk.
Fits of a model exponential velocity field show that the kinematic center of the
molecular gas is consistent with the position of the optical nucleus, within
0.5\asec.  The global CO kinematic position angle, calculated from those fits and
also via the method of \citet{davor}, is $-78^\circ$~\error~3\deg\  to the
receding major axis.  More detailed kinemetric analysis suggests a gradual trend
from PA $-78^\circ$ in the inner resolution element to $-72^\circ$ at radii $>$
6\asec.
In comparison, the stellar photometric major axis is $-67^\circ$ and the stellar
kinematic major axis is $-69^\circ$ \citep{cappellari07}, so the molecular gas (especially at its outer
radii) is aligned with the stellar body of the galaxy to within a few degrees.
Figure \ref{n4526majax}, the major axis position-velocity diagram, clearly
shows that the gas traces a turnover in the rotation curve with
concentrations of gas near the turnover radius.  There is a modest degree of
asymmetry, with somewhat stronger CO emission on the receding (west) side of
the major axis.

\section{Discussion}

\subsection{Origins of the Molecular Gas}

In NGC~4459 and NGC~4526 the molecular gas kinematics are consistent with both the
stellar photometric axes and the stellar kinematic axes.  The differences are on
the order of a few degrees, which is the level of precision available from
existing data.  In NGC~4150 the molecular gas kinematics
are more difficult to describe but the orientation of the dust ring is also
consistent with the stellar photometric and kinematic axes to a degree or better.
In NGC~3032 the molecular gas kinematic position angle is consistent with the
photometric axis to within a few degrees, but it is nearly 180\deg\ away from the
stellar kinematic axis.  Despite the dramatic counterrotation of the gas in NGC
3032, these alignments imply that all four galaxies have nearly axisymmetric
potentials and that the gas is well settled into the equatorial plane.

In NGC~4150, NGC~4459, and NGC~4526 the agreements between the sense of
rotation of the molecular gas and stars
also imply that the molecular gas could have originated
in internal stellar mass loss.  Such an internal origin is not
required by the data, of course, as the gas could have been captured into prograde rotation
from outside sources, and internal secular evolution over several dynamical
timescales would gradually bring it to the galaxy's equatorial plane.
(Orbital periods are a few $10^7$ yr at the edges of the disks.)

In contrast to the other three galaxies,
the conspicuous counterrotation of the molecular gas and stars in NGC~3032 provides clear
evidence that this molecular gas could not have originated in internal stellar
mass loss.  The only scenario that could explain the counterrotation of
internally produced gas is to invoke some gravitational interaction which strongly
torques the gas (possibly at large radii) while leaving the remainder of the
galaxy seemingly untouched.  This scenario seems unnecessarily complex, however,
so we suggest instead that the molecular gas was captured through ``cold"
accretion from the intergalactic medium or in an interaction or a minor
merger with a gas-rich neighbor.  Perhaps it is even a
remnant of a major merger which formed the present galaxy.  The high
degree of regularity in the gas kinematics and stellar morphology
suggests that this event did not occur recently, as
the orbital period at the outer edge of the CO disk is
on the order of $10^8$ yr.  Furthermore, modeling the optical image (Section
\ref{vcirc}) shows that the isophotes are very regular to at least 70\asec\ (7.3
kpc, or five times the radius of the molecular gas).

NGC~3032
is significant, then, because it is one of the few early-type galaxies
in which the possibility of an internal origin for the molecular gas
can be firmly excluded.  It contradicts the picture
outlined by \citet{sw06}, who suggested that most of the molecular gas in
lenticulars should have an internal origin like that described by
\citet{temi07}.
The evidence for the suggestion of \citet{sw06} came from contrasting the atomic and
molecular properties of lenticulars: the atomic gas, which is usually more
extended and shows different kinematics than the molecular gas, tends to
dominate the gas content in gas-rich galaxies, whereas the molecular phase
dominates in gas-poor galaxies.  Thus \citet{sw06} hypothesized
that the atomic gas could be primarily attributed to external sources but
the molecular gas to internal sources.  While that picture may still be
roughly correct, it cannot account for the molecular gas in NGC~3032.

On the other hand, it must still be true that the evolved stars in NGC 3032
expelled mass into the ISM.  According to the models of \citet{FG76} and
\citet{ciotti91}, given the galaxy's present luminosity the stellar mass loss
would amount to 4\e{8} to 6\e{9} \solmass\ over a Hubble time.  We have not
detected that gas yet, as it would show prograde rotation.  The stellar mass
loss must now be in the form of hot gas, or perhaps it has been removed from
the galaxy entirely
\citep[e.g.][]{temi07}.

It is not unusual to find counterrotating ionized gas in lenticular galaxies.
From their own data and a compilation of the literature, \citet{bureau06} find that 15\% \error 4\% of all S0s contain counterrotating
ionized gas while 23\% \error 5\% of the S0s with ionized gas have that gas in
retrograde motion.  Earlier work by \citet{bertola92} and \citet{kuijken96} is
in agreement with this result.
A more detailed study of stellar - ionized gas misalignments was carried out by
\citet{sarzi06} using integral-field observations.  They find that
among the lenticulars of the SAURON early-type sample, 9 of 20 galaxies have
a star-gas kinematic misalignment
greater than 30\deg.  NGC 3032 is the only one of the 9 that
has a misalignment $\ge$
150\deg, though, so it is somewhat unusual in having ionized gas so nearly
retrograde.
We should be careful not to make assumptions about the behavior of the
molecular gas based on the behavior of the ionized gas, however, since in general
it is not obvious that these two phases of the interstellar medium are closely
linked.  For example, in NGC~4150 the ionized gas emission \citep{sarzi06} is
distributed over a much larger scale than the molecular gas; ionized gas
fills the 30\asec~\by~40\asec\ SAURON field of view whereas molecular gas is concentrated in the
central 6\asec.  Clearly, a
larger sample of resolved molecular maps of lenticular galaxies will be
required before we can make better judgments about the origin of their
molecular gas.

\subsection{Deriving circular velocities}
\label{vcirc}

Molecular gas is naturally cold and dissipational and therefore, in equilibrium in an axisymmetric potential, it should settle in circular orbits in the equatorial plane.  To the extent that the velocity dispersion of the molecular gas is low (typically on the order of 10 to 20 \kms) it should then provide an excellent tracer of the circular velocity of the galaxy. And the circular velocity is, ultimately, the dynamical indicator of the total matter content of a galaxy.

It is also possible to estimate the circular velocity from stellar kinematic data, with additional complications. For a given edge-on axisymmetric potential, dimensional arguments suggest that it may be possible to uniquely recover the three-dimensional (3D) orbital distribution of the stars in a galaxy with another 3D quantity, namely the knowledge of the stellar line-of-sight velocity-distribution at every position on the galaxy, as can be obtained with integral-field spectroscopy. It seems however unlikely that the axisymmetric potential itself, which is another two-dimensional function can also be uniquely recovered from the same 3D observations \citep[see Section~3 of][for a discussion]{valluri04}. In the stellar dynamical models either a constant  mass-to-light ratio ($M/L$) or a parametric form for the potential are {\em assumed}. In this context it is worthwhile to compare the two different estimates of circular velocity in early-type galaxies coming from different assumptions.

\citet{cappellari06} have used the stellar kinematics presented in \citet{emsellem04} to derive high quality dynamical $M/L$ for a subset of the SAURON early-type galaxies, under the assumption of a constant $M/L$. High resolution and wide-field images of the galaxies are modeled with a Multi-Gaussian Expansion \citep[MGE;][]{emsellem94,cappellari02} so that, if the inclination of the galaxy (assumed to be axisymmetric) is known, the fitted MGE model of the surface density can be deprojected and a three-dimensional stellar distribution can be estimated. Full two-dimensional maps of the stellar mean velocities, dispersions, and Gauss-Hermite $h_3$--$h_6$ parameters within about an effective radius are then used in combination with the stellar distribution to construct self-consistent two-integral Jeans and three-integral Schwarzschild models.  The data are consistent with the assumption of a constant dynamical $M/L$ within an effective radius, and in galaxies with regular dust disks the inclination inferred from the Jeans models is consistent with the axial ratio of the dust disk.

Here we compute the circular velocities:
\[
v_c^2(R)=R\frac{\partial\Phi(R,z)}{\partial R},
\]
where the gravitational potential $\Phi(R,z)$ is computed from the MGE models tabulated in \citet{cappellari06}, deprojected at their best fitting inclination, and scaled by the constant best-fitting Schwarzschild's $M/L$. For NGC~3032 no previous dynamical model exist so the MGE model, inclination and $M/L$ were determined as described below. The circular velocity curves of the best-fit gravitational potentials for NGC~3032, NGC~4150, NGC~4459 and NGC~4526 are shown in Figures \ref{n3032majax}, \ref{n4150majax}, \ref{n4459majax}, and \ref{n4526majax}.

For NGC~3032 we constructed an MGE model for the photometry using the software of \citet{cappellari02}. To reduce the effect of dust extinction in the derivation of the stellar density distribution, in the MGE fit we combined an archival near-infrared HST/NICMOS/F160W ($H$-band) image of the dusty central regions with a larger-field image taken with HST/WFPC2/F606W ($V$-band). The $V$-band image was adopted as photometric reference and used to converted the MGE model to the Johnson band using the calibration of \citet{dolphin00}. The distance-independent parameters of the PSF-deconvolved MGE models, corrected for galactic extinction following \citet{schlegel98}, and adopting an absolute magnitude of the Sun $M_{\odot,V}=4.83$ mag \citep[Table~2.1 of][]{binney98}, are given in Table~\ref{3032mge}.

We constructed a self-consistent two-integral axisymmetric MGE Jeans model for the second velocity moments $\mu_2\equiv\sqrt{v^2+\sigma^2}$ of NGC~3032 and we determined the best-fitting $(M/L)_V\approx1.7$ and inclination $i\approx42^\circ$, as in \citet{cappellari06}. Using the $V-I=1.07$ color of \citet{tonry01} this translates into $(M/L)_I\approx1.3$, which is among the lowest measured values for any SAURON galaxy \citep{cappellari06}. This is still consistent within the scatter with their $(M/L)-\sigma$ relationship, considering the low value of $\sigma_e=90$ km s$^{-1}$ of NGC~3032 from \citet{emsellem07}. The low dynamical $M/L$ is also expected from their relation between $(M/L)-{\rm H}\beta$ and the fact that NGC~3032 has the lowest average H$\beta$ line-strength among all the galaxies in the SAURON sample of E/S0 galaxies \citep{kuntschner06}. This confirms that the variations in the stellar populations are the main driver for the observed differences in the dynamical $M/L$. However the self-consistent model does not provide a good fit to the data, so the fitted $M/L$ represents an average value for the central regions of the galaxy which are sampled by the SAURON kinematics. In particular the self-consistent two-integral model shows a much steeper radial {\em decrease} of $\mu_2$ along the major axis than observed. This is the opposite of what is generally measured in flattened galaxies and disks \citep[e.g.][]{cappellari07} and suggests that anisotropy cannot explain the discrepancy. An improved non-self-consistent model will be discussed in Section~\ref{ringmodels}.

\begin{deluxetable}{cccc}
\tablewidth{0pt}
\tablecaption{MGE parameters for the deconvolved $V$-band stellar surface brightness of NGC 3032
\label{3032mge}}
\tablehead{
\colhead{$j$} & \colhead{$\log I'_j$} & \colhead{$\log \sigma_j$} & \colhead{$q'_j$} \\
\colhead{} & \colhead{(L$_{\sun,V}$ pc$^{-2}$)} & \colhead{(arcsec)} & \colhead{} }
\startdata
1  &   6.055  &  -1.426  &   0.895 \\
2  &   5.117  &  -0.775  &   0.895 \\
3  &   4.236  &  -0.485  &   0.895 \\
4  &   3.882  &  -0.019  &   0.840 \\
5  &   3.005  &   0.282  &   0.950 \\
6  &   2.797  &   0.597  &   0.863 \\
7  &   1.769  &   0.913  &   0.950 \\
8  &   2.060  &   1.098  &   0.800 \\
9  &   1.635  &   1.532  &   0.806 \\
\enddata
\tablecomments{$I'_j$ is the peak surface brightness
of the $j$-th Gaussian, which has dispersion $\sigma_j$
and observed axial ratio $q'_j$.}
\end{deluxetable}

\subsection{Comparing circular velocities with the observed CO velocities}\label{ringmodels}

We make a comparison between the stellar and the molecular indicators of circular velocity with tilted ring models, projected and convolved to the same resolution and sampling as the CO data. The GALMOD routine in the GIPSY software package is used to generate the models.  We specify the rotation velocities of the rings as computed in Section~\ref{vcirc} and listed in Tables \ref{4526model}, \ref{4459model}, and \ref{3032model}. The inclinations of the gas disks are also known, as described above, and the velocity dispersion in the molecular gas is assumed to be 10 \kms. The exact value of the velocity dispersion used is not critical, as long as it is smaller than the channel width and consistent with the sharp cutoffs in emission at the outlying velocities. Nominally the rotation velocities should also be corrected for this finite velocity dispersion (the asymmetric drift effect), but as the rotation velocities are 100 to 350 \kms\ the correction is negligible. The optical dust images also suggest a very thin molecular disk, here assigned a scale height of 1\asec\ (again the exact value is not critical as it is smaller than the spatial resolution). A gradual linear decrease in the gas surface density is assumed, with a sharp cutoff at the edge of the dust disk. The rotating disk model is then projected onto the plane of the sky at the specified inclination and the model emission is sampled at a pixel size and channel width matching the CO observations.  The sampled cube is also spatially smoothed to have the same resolution as the CO beam. Finally the smoothed model cube is sliced along the galaxy major axis just as for the real data, and the slice is displayed with analogous contour levels (as percentages of the peak intensity, which is arbitrary).

\begin{figure}
\includegraphics[scale=0.7]{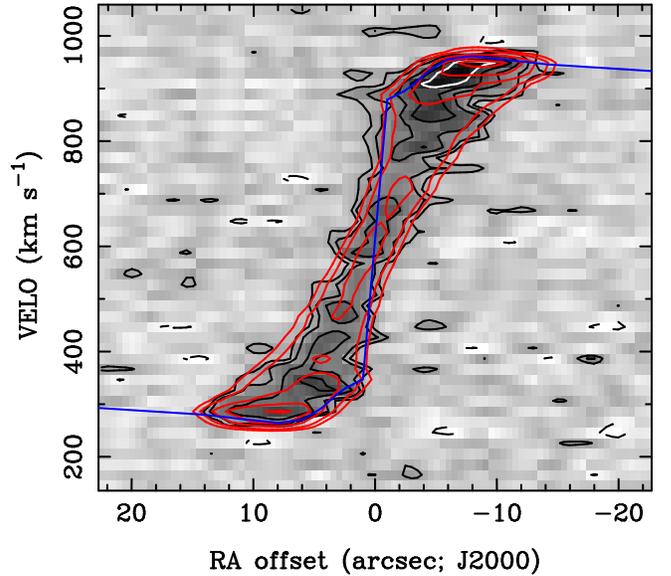}
\caption{Observed and modeled position-velocity diagram for NGC 4526.  The
greyscale, black, and white contours show the CO intensity as in Fig.~\ref{n4526majax};
red contours show the corresponding position-velocity diagram for the tilted
ring model in which the circular velocity is everywhere 5\% smaller than that given by
\citet{cappellari07}, as discussed in Section~\ref{ringmodels}. The adjusted $V_c\:\sin{i}$ are shown in the blue line.  Contour levels for the model
position-velocity diagram are at the same levels (relative to the peak intensity)
as for the data.
\label{4526model6}
}
\end{figure}

Figure \ref{4526model6} compares the data and the model position-velocity diagrams for NGC~4526. 
The assumed gas surface densities drop linearly by a factor of 5 between 1\asec\ and 14\asec\ and are zero beyond that. 
Placing the gas in rings rotating at the circular velocity produced a model with
a bit too large velocity amplitude; the
peaks in the model position-velocity diagram, tracing gas along the line of nodes, 
were separated by 703 \kms\ whereas in the data they are separated by 
663 \kms\  (the difference is two channels).
Therefore, Figure \ref{4526model6} actually shows a model in which the
rotation velocity of each ring is 5\% smaller than the circular velocities derived with the $M/L$ of \citet{cappellari06}, and it shows that this model 
produces a very close match to
both the ridgeline and the envelope of the observed emission. 
Even the rounded shoulder of the circular velocity curve (specifically, the relatively shallow but still rising slope in $1'' \le r \le 8''$) is confirmed by the CO data. 
We have not made an exhaustive search in the rotation velocity or gas distribution parameter space, so we certainly cannot claim to have made the unique best match to the observed position-velocity diagram.  
However, the good agreement between the model and the data should be regarded as independent confirmation that the circular velocities derived with the $M/L$ of \citet{cappellari06} are accurate to about 5\%.  
The $M/L$ is thus accurate to about 10\%, roughly consistent with the quoted uncertainty.

\begin{figure}
\includegraphics[scale=0.6]{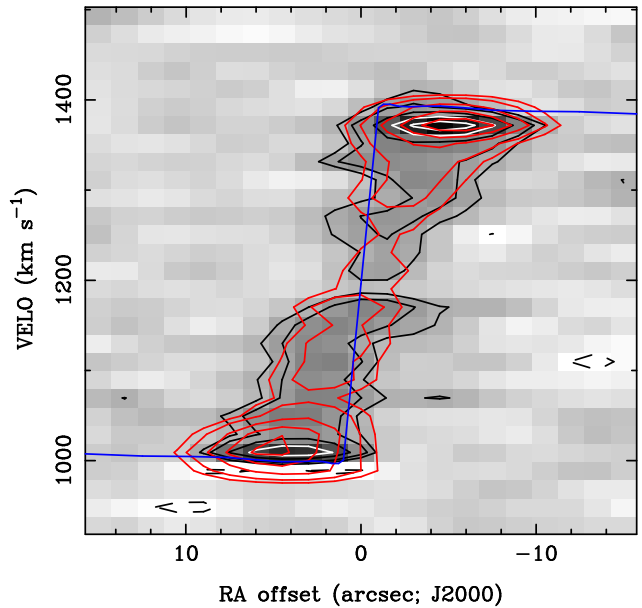}
\caption{Observed and modeled position-velocity diagram for NGC 4459.  The
greyscale, black, and white contours show the CO intensity as in Fig.~\ref{n4459majax};
red contours show the model position-velocity diagram in which the rotation
velocity of each ring is the circular velocity of \citet{cappellari06} (blue line).
\label{4459model2}
}
\end{figure}

Figure \ref{4459model2} presents a similar comparison for NGC~4459.  In this case the circular velocities obtained with the $M/L$ of \citet{cappellari06} are used without correction and an excellent match is achieved, indicating again the high degree of accuracy of the dynamical $M/L$. Figures \ref{n4459majax} and \ref{n4526majax} also show that in NGC~4459 and NGC~4526 the molecular gas rotation speeds are 1.6 to 2.0 times larger than the mean stellar velocities at the same radii.  In these galaxies the stars have significant pressure support, but the molecular kinematics and the stellar kinematics are both consistent with the same gravitational potential. 

\begin{deluxetable}{lccc}
\tablewidth{0pt}
\tablecaption{Tilted Ring Models for NGC 4526
\label{4526model}}
\tablehead{
\colhead{Radius} & \colhead{$V_{rot}(1)$} & \colhead{$V_{rot}(2)$} & \colhead{$\Sigma_{gas}$} \\
\colhead{\asec} & \colhead{\kms} & \colhead{\kms} & \colhead{} }
\startdata
1.34 & 289.0 & 274.6 & 1.00 \\
1.80 & 295.2 & 280.4 & 0.97 \\
2.42 & 302.7 & 287.6 & 0.93 \\
3.26 & 318.0 & 302.1 & 0.88 \\
4.37 & 343.3 & 326.1 & 0.81 \\
5.88 & 366.4 & 348.1 & 0.72 \\
7.90 & 373.8 & 355.1 & 0.59 \\
10.62 & 366.6  & 348.3 & 0.43 \\
14.26 & 356.8 & 339.0 & 0.20 \\
\enddata
\tablecomments{The velocities $V_{rot(1)}$ are equal to the stellar-derived
circular velocities, and $V_{rot(2)} = 0.95 V_{rot(1)}$.}
\end{deluxetable}

\begin{deluxetable}{lcc}
\tablewidth{0pt}
\tablecaption{Tilted Ring Model for NGC 4459
\label{4459model}}
\tablehead{
\colhead{Radius} & \colhead{$V_{rot}$} & \colhead{$\Sigma_{gas}$} \\
\colhead{\asec} & \colhead{\kms} & \colhead{} }
\startdata
1.5 & 272.3 & 1.00 \\
3.0 & 268.7 & 0.86 \\
4.5 & 269.6 & 0.73 \\
6.0 & 266.8 & 0.60 \\
7.5 & 263.3 & 0.47 \\
9.0 & 262.1 & 0.10 \\
\enddata
\end{deluxetable}

A broader implication of this agreement is that the molecular and stellar 
kinematics for galaxies like NGC~4459 and NGC~4526 can be used with confidence 
in studies of the circular velocity and the bulge velocity dispersion 
\citep[e.g.][]{courteau07}, where it is necessary to probe the relationships
between the two dynamical indicators.  For example,
several authors \citep[e.g.][]{ho07,shields06} have studied the
development of the relationship between a galaxy's black hole mass and 
its bulge velocity dispersion.  They have advocated use of the CO line width to
trace the galaxy's maximum circular velocity, hence its velocity dispersion
(especially at high redshift where it is currently easier to measure the CO line
width than other dynamical indicators of the galaxy mass).  These studies use a
CO Tully-Fisher relation to demonstrate that the CO line widths do indeed trace
their host galaxies' maximum circular velocity.  Our detailed comparison of
molecular kinematics and circular velocities supports this claim when
the CO velocities and circular velocities at large radii are considered, even
for early-type galaxies where the Tully-Fisher relation is not traditionally
applied.

Figure \ref{n4150majax} compares the CO position-velocity diagram in NGC~4150 to the inferred circular velocity.  In this case the total velocity range covered by the molecular gas is nearly as large as what would be expected from the circular velocity curve.  The CO velocities are also nearly equal to the mean stellar velocities at radii $\gtrsim$ 10\asec, which is qualitatively consistent with the fact that the observed stellar velocity dispersions are much smaller than in NGC~4459 and NGC~4526 \citep{emsellem04}. It is worth noting that the adopted inclination for the galaxy, while consistent with the axis ratio of the dust ring, may not be appropriate for the molecular gas which is concentrated in the nuclear dust lane. Detailed models are probably not yet useful for these data, and the most that can be said is that the inferred circular velocity curve is not inconsistent with the observed CO and stellar velocities.

In contrast to the cases of NGC~4459 and NGC~4526, there are significant disagreements between the inferred circular velocity and the observed molecular gas rotation speeds of NGC~3032. Assuming a constant dynamical $M/L$ within an effective radius, the circular velocity rises quickly to a peak of 330 \kms\ at 0.08\asec\ and makes a sharp, nearly Keplerian decline through the inner arcseconds.  It drops by a factor of two to 157 \kms\ at 5\asec\ and thereafter drops more slowly, passing through 132 \kms\ at 10\asec\ which is nearly the edge of the molecular disk.  But as Figure \ref{n3032majax} shows, these velocities are much larger than the observed CO velocities especially interior to 5\asec.

\begin{figure}
\includegraphics[scale=0.5]{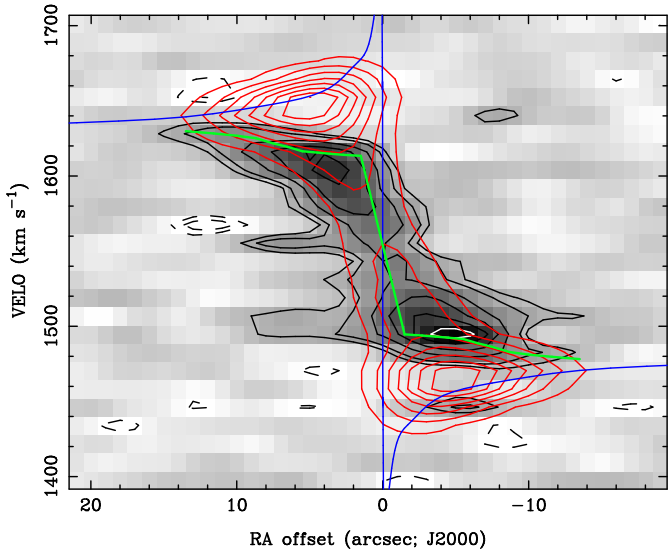}
\caption{Observed and modeled position-velocity diagram for NGC 3032.  The
greyscale, black, and white contours show the CO intensity as in
Fig.~\ref{n3032majax};
red contours show a model position-velocity diagram in which the rotation
velocity of each ring is the circular velocity (blue line), assuming a decrease of a
factor of two in the M/L interior to 10\asec.
The green line shows a rotation curve $V_{rot}\:\sin{i}$ which reproduces the
observed gas velocities, as in Table \ref{3032model}.
\label{3032model2}
}
\end{figure}

The strong radial gradient in the \hbeta\ absorption line strength index of
NGC~3032 \citep{kuntschner06}, and the young inferred central age
\citep{mcdermid06a} suggest that the stellar population might be better described with a variable $M/L$. 
Indeed, if the line strength indices are interpreted with the models of \citet{thomas03} and \citet{maraston05}, the SAURON data are consistent with a decline of a factor of two to three in the local stellar $M/L$ from 10\asec\ to 1\asec, even in the $H$ band (from which the MGE models are derived). 
Moreover the inability of the self-consistent Jeans model of Section~\ref{vcirc} to qualitatively reproduce the observed kinematics also indicates the need for a varying dynamical (total) $M/L$. 
Thus a second MGE model was constructed, with a variable $M/L$, such that the deprojected Gaussian components having widths (dispersions) $\leq$ 1\asec\ have M/L values half as large as the components with dispersions $\geq$ 10\asec, and the variation is logarithmic between those extremes. 
Using this mass model a non-self-consistent Jeans model was computed and again fitted to the observed SAURON stellar kinematics to derive a new $M/L$. 
Interestingly this revised Jeans model, which has an $M/L$ gradient dictated by the stellar population measurements, now also well reproduces the observed SAURON stellar kinematics. 
The inferred circular velocity curve has a somewhat smaller peak velocity, 285 \kms\ at 0.08\asec, a steeper decline in the inner few arcseconds and a shallower slope at 10\asec, but its amplitude is the same at 10\asec\ (132 \kms) as for the model with the constant $M/L$.  
Figure \ref{3032model2} shows the predicted CO velocities that would have been observed if the molecular gas in NGC~3032 were rotating at the inferred circular velocity (Table \ref{3032model}).  
Clearly the assumption of a variable M/L does not remove the discrepancy between observed and expected velocities.

A reasonably accurate measure of the discrepancy between observed and expected velocities can be made from a model which describes the CO velocities well.  
Table \ref{3032model} and Figure \ref{3032model2} show the rotation velocities in
this model; they have been altered by hand to reproduce the observed position-velocity diagram. 
As before, no formal fitting has been done so the model shown cannot be claimed unique or best and its gross properties should be trusted to a greater degree than its details.  
This ``working" model has a rotation velocity of 98 \kms\ at 6.5\asec\ and 113 \kms\ at 13.5\asec.  
At the outer edge of the CO disk its velocity is only 10\% smaller than that of the inferred circular velocity curve.  
However, at 6.5\asec\ (700 pc) its rotation velocity is only 70\% of the circular velocity curve, which means that the enclosed mass is a factor of two smaller than that suggested by the circular velocity.  
The discrepancy is even worse at smaller radii, where the circular velocity is nearly a factor of three higher than the rotation velocities in the molecular gas and the implied mass difference is a factor of 10.

\begin{deluxetable}{lccc}
\tablewidth{0pt}
\tablecaption{Tilted Ring Models for NGC 3032
\label{3032model}}
\tablehead{
\colhead{Radius} & \colhead{$V_{circ}$} & \colhead{$V_{rot}$} & \colhead{$\Sigma_{gas}$} \\
\colhead{\asec} & \colhead{\kms} & \colhead{\kms} & \colhead{} }
\startdata
1.5 & 183.6 & 88.7 & 1.0   \\
2.7 & 166.1 & 89.8 & 0.9   \\
3.6 & 155.3 & 90.9 & 0.8   \\
4.5 & 148.8 & 92.0 & 0.7   \\
5.5 & 144.5 & 93.2 & 0.6   \\
6.5 & 141.4 & 97.6 & 0.5   \\
7.5 & 138.7 & 101.0 & 0.4  \\
8.7 & 135.7 & 106.3 & 0.3  \\
10.0 & 132.4 & 108.8 & 0.2 \\
11.6 & 129.1 & 111.0 & 0.1 \\
13.5 & 126.1 & 113.3 & 0.0 \\
\enddata
\tablecomments{The circular velocities are derived under an assumption of a
variable stellar mass-to-light ratio, as described in the text.  The CO
rotation model provides a reasonably good (though fit by eye)
match to the observed position-velocity diagram.
In both cases the gas surface density is taken to be
zero interior to 1.5\asec.}
\end{deluxetable}

At present the source of the discrepancy between the observed CO velocities
and the inferred circular velocity curve is not understood. The unusually
strong peak in the circular velocity curve is driven by the very bright blue
nucleus in optical images and by the high central stellar velocity
dispersion. Therefore, one possibility is that the measured stellar mean
velocity and/or velocity dispersion in the center of NGC~3032 are
significantly overestimated. However, the velocity dispersion measurements in
the central regions made by \citet{mcdermid06a} with the OASIS instrument are consistent with the SAURON data. Another striking possibility is that the molecular gas in NGC~3032 may not be rotating at the circular velocity.  We argued above that the gas had to have been acquired from some external source, and perhaps it is still undergoing strong dynamical evolution as it settles into equilibrium. Yet if its velocities are $\leq$ 70\% of the circular velocity its orbits would be highly noncircular, which seems inconsistent with the regular, symmetric appearance of the CO velocity field (Figure \ref{n3032vel}).  Neither of these explanations is completely satisfactory.

In short, the CO velocities in NGC~4459 and NGC~4526 are in excellent agreement
with the circular velocities inferred for those two galaxies.  These matches
indicate that the stellar and CO kinematics are well understood. 
In NGC~3032 and NGC~4150 at radii $\gtrsim$ 10\asec, the observed CO velocities are nearly equal to the mean stellar velocities and 10\% to 20\% smaller than the inferred circular velocities. 
However, the CO velocities of NGC~3032 are conspicuously low when compared to its inferred circular velocity within 6\asec\ of the nucleus, and it would be valuable to carry out this kind of a comparison in other galaxies with counterrotating molecular gas in order to ascertain whether NGC~3032 is unusual.

\section{Molecular and Ionized Gas Velocities}

In normal spiral galaxies, the bulk of the ionized gas emission traces HII
regions recently formed out of the molecular gas and sharing the disky kinematics of
the molecular gas.  The effect is that in a major axis position-velocity
diagram the ionized gas tends to trace the ridgeline of the molecular and
atomic gas or to be displaced towards the high velocity envelope (due
primarily to resolution effects).  Examples can be seen in \citet{kregel} and
\citet{ngc4647}.  Thus, comparisons between the molecular and ionized
kinematics can help to elucidate the role of star formation activity (or the
lack thereof) in producing the ionized gas.

Excluding the case of NGC~4150, which is discussed in \SS\ref{4150results},
there appears to be a systematic difference between ionized gas kinematics
and molecular kinematics of these lenticulars.  The rotation speeds of the
ionized gas are $\sim$  20\% smaller than those of the CO.
There is also a slight tendency for the \oiii\ velocities to be smaller than
\hbeta\ velocities; this latter feature is most obvious at radii $>$ 5\asec\
in NGC~4526 (Figure \ref{n4526majax}).
Lower velocities in the ionized gas than in the molecular gas
is the opposite situation to what is usually observed in the disks of spirals and is
also in the opposite sense to the effects of beam smearing on the CO
velocities.  These data suggest that there may be a kinematic component of the
ionized gas which is not dynamically cold and not related to star
formation.  They also imply that if the ionized gas in early-type galaxies is
to be used for a measurement of the galaxy's circular velocity, an asymmetric
drift correction may be significant \citep[e.g.][]{cretton00}.
By way of a caveat, though, we note that
these lenticulars are dusty and
NGC~4526 in particular has a high inclination (79\deg), so the observed CO
and ionized gas velocities may be biased by projection and optical depth effects.
The details of the relationships between cold and warm ionized phases of the
interstellar medium are not yet clear, and they deserve closer scrutiny for
the insights they may give into the evolution of the ISM in early-type
galaxies.

\section{Summary}

We present resolved images of the CO emission in the four lenticular galaxies
NGC~3032, NGC~4150, NGC~4459, and NGC~4526.  These are some of the most
CO-rich galaxies in the SAURON survey of early type galaxies, so they are
prime targets for investigations which use cold gas to trace the
interaction/merger history of early type galaxies and also to document
morphological change through star formation and disk growth.  Their inferred
\htoo\ masses are in the range
5\e{7} to 5\e{8} \solmass.  The molecular gas
is located in kpc-scale disks (in excellent agreement with the distribution
of dust visible in broadband optical images), the smallest being NGC 4150
with a radius of 500 pc and the largest being NGC 3032 with a radius of 1.5
kpc.  Average molecular surface densities (including helium) are 100 to 200 \msunsqpc.

In three of the four galaxies (NGC 3032, NGC 4459, and NGC 4526) the molecular gas is distributed in
disks which show regular rotation and little sign of recent disturbance.  The
kinematic major axes are well aligned with the host galaxies' photometric and
(stellar) kinematic major axes, suggesting that the gas has settled into the
equatorial plane of nearly axisymmetric potentials.  The
velocity field of NGC 4150 shows a kinematic major axis in rough agreement
with the galaxy's optical major axis, but better spatial resolution will be
necessary in order to assess the regularity of the CO kinematics.
Furthermore, in NGC 3032 the molecular gas's kinematic position angle is
180\deg\ offset from the stellar kinematic position angle; this dramatic
counterrotation indicates that the molecular gas was acquired from an
external source or perhaps is leftover from a major merger, but it cannot
have been produced through internal stellar mass loss.
The sense of the CO rotation in NGC~3032 is consistent with that of the young
kinematically decoupled core, however, which suggests that this is an example
of a stellar substructure forming through dissipational processes.

In two cases (NGC 4459 and NGC 4526) the CO kinematics provide powerful,
independent confirmation of the dynamical mass-to-light ratios inferred by
\citet{cappellari06}.  The mass-to-light ratios were derived from the SAURON
stellar kinematic data via two-integral Jeans and three-integral
Schwarzschild dynamical models and were used to infer circular velocity
curves.  Simple tilted ring models are presented in which the molecular gas
rotates at the circular velocity (or, in NGC 4526, 95\% of the circular
velocity, a decrease which is roughly consistent with the uncertainty in the mass-to-light ratio).
Comparison of the model and observed major axis
position-velocity diagrams indicates that the circular velocity inferred from
stellar kinematics alone is consistent with the behavior of the molecular gas.
The agreement is significant because it is rare to be able to make this kind
of an independent check of the stellar dynamical analysis.

Several puzzling results of this work deserve further study; these are
disagreements between the gaseous and the stellar kinematics at sub-kpc
scales.
For example, in NGC~4150 the CO velocities
appear to be inconsistent with those of the young (counterrotating) stellar
core.
In NGC 3032 the observed CO velocities are
significantly lower than the circular velocity curve (amounting to an implied
factor of two less mass interior to 700 pc and even more at smaller radii).
At present it is not known whether the stellar kinematic data give
an overestimated circular velocity for this galaxy or whether the recently
accreted molecular gas is not following circular orbits.
Higher resolution CO observations will also be necessary to check
whether NGC~4150 shows an inconsistency between CO velocities and circular
velocities.  Finally, the relationships between molecular gas and ionized gas should
contribute valuable insights into star formation and the evolution of
the ISM in these lenticulars.

\acknowledgments

Davor Krajnovi\'c kindly provided his IDL routines for kinemetric analysis and
assistance in using them.
LMY thanks the University of Oxford sub-department of Astrophysics for
hospitality during a sabbatical visit and
acknowledges support from NSF AST-0507432. MC acknowledges support from a PPARC Advanced Fellowship (PP/D005574/1).
This work is partially based on observations made with the NASA/ESA Hubble
Space Telescope, obtained from the data archive at the Space Telescope
Science Institute.  STScI is operated by the Association of Universities for
Research in Astronomy, Inc.\ under the NASA contract NAS 5-26555.

Facilities: \facility{BIMA}



\end{document}